\documentclass[12pt,preprint]{emulateapj}
\usepackage{amsmath}
\usepackage{rotating}
\begin{document}
 
\pagestyle{plain}
\newcommand{\Msun}{M$_\odot$}
\newcommand{\Lsun}{L$_\odot$}
\newcommand{\Zsun}{Z$_\odot$}
\newcommand{\degree}{\ensuremath{^\circ}}
\newcommand{\cm}{cm$^{-3}$}
\newcommand{\na}{NewA}

\title{Kinetic energy from supernova feedback in high-resolution galaxy simulations}
\author{Christine M. Simpson$^{1}$, Greg L. Bryan$^{2}$, Cameron Hummels$^{3}$, Jeremiah P. Ostriker$^{2}$}

\altaffiltext{1}{Heidelberger Institut f\"{u}r Theoretische Studien, Schloss-Wolfsbrunnenweg 35, 69118 Heidelberg, Germany; Christine.Simpson@h-its.org}
\altaffiltext{2}{Department of Astronomy \& Astrophysics, Columbia University, New York, NY 10027 USA}
\altaffiltext{3}{Department of Astronomy and Steward Observatory, University of Arizona, Tucson, AZ 85721, USA}

\begin{abstract}
We describe a new method for adding a prescribed amount of kinetic energy to simulated gas modeled on a cartesian grid by directly altering grid cells' mass and velocity in a distributed fashion.  
The method is explored in the context of supernova feedback in high-resolution ($\sim 10$ pc) hydrodynamic simulations of galaxy formation.  Resolution-dependence is a primary consideration in our application of the method and simulations of isolated explosions (performed at different resolutions) motivate a resolution-dependent scaling for the injected fraction of kinetic energy that we apply in cosmological simulations of a $10^9$ \Msun dwarf halo.    
We find that in high density media ($\gtrsim$ 50 cm$^{-3}$) with coarse resolution ($\gtrsim 4$ pc per cell), results are sensitive to the initial kinetic energy fraction due to early and rapid cooling.  In our galaxy simulations, the deposition of small amounts of supernova energy in kinetic form (as little as 1\%) has a dramatic impact on the evolution of the system, resulting in an order of magnitude suppression of stellar mass.  The overall behavior of the galaxy in the two highest resolution simulations we perform appears to converge.  We discuss the resulting distribution of stellar metallicities, an observable sensitive to galactic wind properties, and find that while the new method demonstrates increased agreement with observed systems, significant discrepancies remain, likely due to simplistic assumptions that neglect contributions from Type Ia supernovae and stellar winds.

\end{abstract}

\section{Introduction}
Stellar feedback is thought to be an important driver in shaping galaxy properties, but it is one of the most difficult effects to model in numerical simulations of galaxy formation.  The channels for feedback from individual stars are generally well understood, but the spatial and mass resolutions typical of galaxy simulations require `sub-grid' modeling of these effects based on assumptions of how feedback energy couples to the interstellar medium (ISM).  Assumptions also have to be made about the number and type of feedback events generated by star particles in the calculation, which are usually massive, collisionless particles representing a well-sampled stellar population.  Models of this type are often tuned to reproduce certain galaxy properties, while treating other galaxy properties as simulation predictions \citep[e.g.][]{marinacci14}.

This difficulty has fundamental implications for the ability of numerical simulations to inform problems in galaxy formation in which stellar feedback is thought to play a crucial role.  Stellar feedback has been indicated as a potential solution to many outstanding problems in galaxy formation in the $\Lambda$CDM paradigm such as the inner density profiles of galaxies \citep{governato12}; the low baryon fractions of galaxies \citep{hopkins13};  and missing metals in galaxies \citep{hummels13}.  It is challenging to interpret simulation results on these questions and others when different sub-grid feedback prescriptions can produce different galaxy properties \citep{sales10,schaye10}.

One of the main challenges encountered in simulating stellar feedback in galaxy simulations is the efficient cooling of thermal feedback energy \citep{Katz:1992p1037, Katz:1996p1031, Steinmetz:1995p1005}.  This high efficiency results from the large masses contained in the individual gas resolution elements (either particle or grid cell) to which feedback energy is added.  These large masses result in a lower specific heating rate and therefore a lower temperature.  For thermal feedback energy to have a significant impact on gas dynamics, it needs to reach temperatures of $\sim 10^6$ K where radiative cooling is less effective and large pressure gradients will cause gas to move \citep{DallaVecchia2012}.  One solution to this problem is to find some way to keep the temperature of feedback-heated gas elevated by restricting radiative cooling in star-forming regions \citep[e.g.][]{Gerritsen:1997p1039, Thacker:2000p1040, SommerLarsen:2003p1116, stinson06, Governato:2007p1022, Agertz:2010p461, Colin:2010p1053, Piontek:2011p1041, guedes11,hummels12}.  A different approach that also prevents large radiative losses is to delay the addition of feedback to the gas until enough energy has accumulated so that the resulting gas temperature is high enough that the radiative cooling time is long \citep[e.g.,][]{Kay2003, DallaVecchia2012}.  Another approach that can decrease radiative losses is to explicitly model a sub-grid multi-phase model \citep{Scannapieco:2006p1118, Marri2003}.

As an alternate to thermal feedback, a number of works have explored kinetic feedback by imparting momentum to particles, both in SPH codes \citep[e.g.,][]{springel03, oppenheimer06, oppenheimer08, DallaVecchia2008} and in AMR codes \citep[with debris particles,][]{dubois08}.  While most models have focused on supernovae feedback in either kinetic or thermal form, stars produce energy in other forms.  For example \citet{Hopkins2011, Hopkins2012a} and later \citet{agertz13} directly modeled the momentum input of stellar radiation using approximate methods for radiative transfer.  While radiative transfer schemes are typically computationally expensive, \citet{Wise2012} included this effect in a full ray-tracing code for low mass systems at high redshift.  \citet{Stinson2013} assumed that 10\% of this (UV) radiation was thermalized and added it directly to the gas thermal energy.  Recent models have also explored the impact of adding feedback from  cosmic rays \citep[e.g.,][]{Jubelgas2008, Salem2014} and local photoionizing sources \citep{Kannan2014}.  The physical motivations for these approaches vary depending on the resolution of the simulation and on the assumptions about the ISM processes being simulated.  

In the real ISM, the process of supernova feedback is mediated by the evolution of individual supernova remnants.  The kinetic energy fraction in real supernova remnants never dips below 28\%.  Initially, 100\% of the energy is in kinetic form, during the free expansion phase of its evolution, when the material ejected from the progenitor moves ballistically outward.  Once the mass of swept-up ISM becomes equal to the ejected mass and the reverse shock has re-heated the remnant interior, the remnant enters the Sedov-Taylor phase which is energy conserving due to the very long cooling time of the post-shock material.  In this phase, the kinetic energy fraction equilibrates to 28\%.  As the remnant expands further, cooling begins to have an effect and the remnant loses thermal energy, but the expansion continues to be driven by the interior pressure, in the pressure-driven snowplow phase.  Eventually, this pressure becomes negligible to the dynamics of the remnant and the expanding shell is carried outward by its own momentum in the momentum-conserving snowplow phase.  During the two snowplow phases the overall energy of the remnant drops due to cooling, however, since this energy is lost from the thermal component, the overall fraction of kinetic energy does not decrease.  Kinetic energy clearly plays an important role in this feedback channel.

Cosmological simulations of low mass galactic systems have recently begun to reach resolutions of order a few parsecs where this process may begin to be resolved \citep{hopkins13}.  While increased resolution does appear to alleviate the overcooling problem to some degree, purely thermal feedback schemes still have problems reproducing some galaxy properties \citep{simpson13}.  High-resolution simulations of supernova remnants predict that, depending on ISM conditions, remnants should still be in an energy conserving Sedov-Taylor phase on these scales \citep{chevalier74} in which approximately 28\% of the supernova energy is in kinetic form.  Careful, one-dimensional studies of supernovae explosions have established that supernova remnants enter a phase of rapid cooling once the remnant reaches a radius of $\sim$20 pc, depending on the energy of the explosion, the density of the media and the cooling model employed \citep{cioffi88}.  Recent three-dimensional simulations (contemporaneous with the writing of this paper) have explored the effect of resolved supernova explosions that include momentum injection in inhomogeneous media with the aim to better understand the properties of the turbulent ISM \citep{walch14,kim15,martizzi15}.

The goal of this paper is to present a new method for the direct addition of kinetic energy to gas simulated on a cartesian mesh and to explore the effect of direct kinetic energy injection with this method in a galactic context.  An important consideration in applying this method is its resolution dependence.  We explore this issue in detail in simulations of isolated explosions where we can control the background density and the fraction of injected kinetic energy.  These idealized simulations are employed to motivate a resolution-dependent scaling for the injection of kinetic energy.  This resolution-dependent scaling is applied to supernova feedback in simulations of a dwarf galaxy where we can explore the impact of our kinetic energy injection method in a more realistic galactic context.  We apply the method to a feedback channel most like Type-II supernovae; however, contributions from Type Ia supernovae and `runaway' OB stars likely play an important role, therefore the goal of these simulations is to isolate the impact of this method for energy injection in a simplified set-up.

The outline of this paper is as follows.  In Section \ref{sec:kin_algorithm} we describe the method for momentum injection as implemented in the adaptive mesh refinement (AMR) code Enzo.  Then, in Section \ref{sec:idealized_tests}, we describe a set of idealized tests of our model focusing first on its behavior at high resolution (Section \ref{sec:high_res_ref}) and second on the resolution dependence of its behavior (Section \ref{sec:res_study}).  We also describe a resolution-dependent scaling for the initial fraction of kinetic energy injected (Section \ref{sec:kinf_var}).  In Section \ref{sec:galaxy_simulation} we present a set of cosmological dwarf galaxy simulations utilizing the new model and explore how the injection of kinetic supernova feedback energy affects the star formation rate and metal enrichment in the simulations.  Finally, we place our model in context with other feedback models and discuss how to apply our results to future simulations (Section~\ref{sec:discussion}).

\section{Method for Kinetic Feedback}

\label{sec:kin_algorithm}

\begin{figure}
\centering 
\includegraphics[scale=0.75] {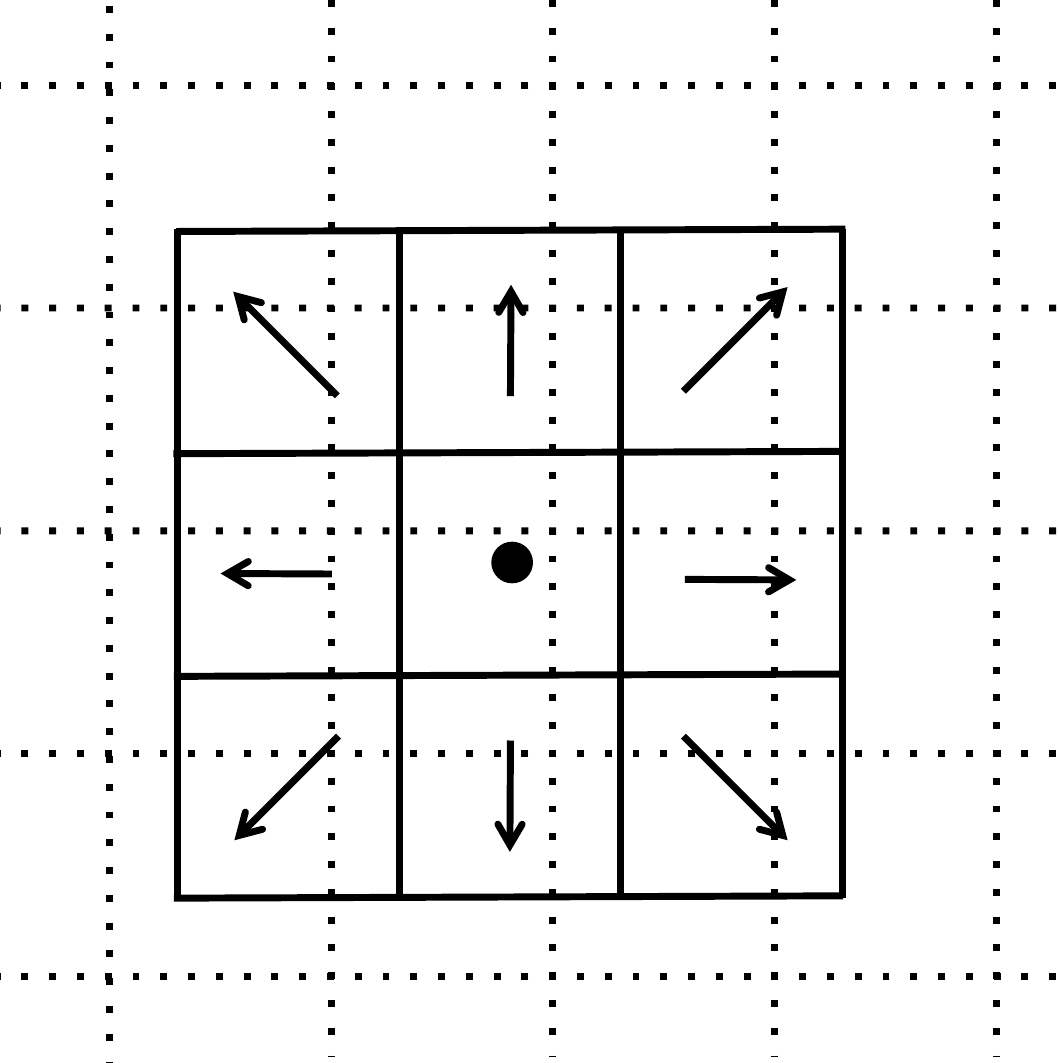}
\caption{Two-dimensional schematic of the virtual cloud stencil (solid lines) relative to the simulation grid (dotted lines) around an individual star particle (black circle).  The direction of the momentum deposition in each virtual cloud cell is shown with an arrow.}
\label{fig:cic}
\end{figure}

In this section, we describe a method for the injection of kinetic supernova energy on a cartesian mesh as implemented  in the AMR code Enzo \citep{bryan14}.  The method we describe is quite general and could be expanded to non-cartesian grids.  The idea is to isotropically add momentum in a distributed region based on the cloud-in-cell (CIC) approach \citep{hockney88}, such that the resulting kinetic and thermal energies are increased by a specified amount.  It differs from previous supernova feedback methods implemented in Enzo that inject purely thermal energy \citep[e.g.][]{tasker06} and from methods implemented in other AMR codes that add an effective pressure term to Euler's equations \citep[e.g.,][]{agertz13}.  The method described here is scale independent in the sense that it describes an algorithm for momentum injection that relies on a feedback zone that is fixed in the number of resolution elements it contains, but not in the physical size of those elements.  For example, this scheme could be adapted to a subgrid treatment for AGN winds in simulations with coarse spatial resolutions of order a kiloparsec.  In this paper, we focus on its use in simulating supernova feedback in galaxy simulations where the spatial resolution is typically no more than 10pc and where at the highest resolutions we consider ($<$ 1pc), the Sedov-Taylor blast region is resolved.  Therefore, rather than conceiving of the strength of the mechanical momentum as a wind speed as done in other treatments of supernova momentum \citep{hopkins13}, we consider the fraction of the supernova energy budget that goes into momentum as an input and the resulting wind speeds as a simulation output. 

The total supernova energy is defined to be a fraction ($\epsilon_{\rm{SN}}$) of the rest mass energy of the star particle producing the feedback ($E_{\rm{SN}} =\epsilon_{\rm{SN}}m_*c^2$, where $m_*$ is the mass of the star particle and $c$ is the speed of light).  In this model, the fraction of the supernova energy that is initially injected as kinetic energy is a free parameter, $f_{\rm{kin}}$, between zero and one.  The kinetic and thermal energies after the supernovae are therefore

 \begin{equation}
\label{eq:kinetic_energy}
E^{a}_{\rm{kin}} = E^b_{\rm{kin}} + f_{\rm{kin}}E_{\rm{SN}}
\end{equation}
 \begin{equation}
\label{eq:thermal_energy}
E^a_{\rm{therm}} = E^b_{\rm{therm}} + (1 - f_{\rm{kin}})E_{\rm{SN}},
\end{equation}

\noindent where $E^b_{\rm{therm}}$ and $E^b_{\rm{kin}}$ are the total thermal and kinetic energies of the feedback injection zone before the energy deposition.

Energy, momentum and mass are added to the grid on a cell by cell basis; however, the momentum needs to be added in such a way that the correct amount of total kinetic energy is added, but the net (vector) momentum added is zero.  
We begin by computing the momentum field $\mathbf{p}^{b}$ in the frame of the moving particle, where the particle velocity is $\mathbf{v}_p$ and the velocity of grid cell $i$ is $\mathbf{v}^b_i$,  so that $\mathbf{p}^{b}_{i} = m^{b}_{i} (\mathbf{v}^{b}_{i} - \mathbf{v}_p)$, where $m^{b}_{i}$ is the mass of the cell.  Once the momentum-updating procedure described below has been completed, the momentum can be converted back to velocity in the original frame of the simulation grid.

To add mass, momentum and thermal energy to the comoving grid, we adapt the well-known CIC method \citep{hockney88}.  To remind readers, the CIC method is a way of smoothly adding particle mass to a grid.  Conceptually, this is done by virtually converting the particle into a uniform `cloud' with a size equal to one cell width.  Since the particle position is generally not coincident with a cell center, this region will overlap 8 cells on the grid in three dimensions.  The amount of mass added to each cell is proportional to the volume overlap of the virtual cloud with that cell.

We adopt this same idea, but instead of using a region of size 1$\times$1$\times$1 cell widths, we use a 3$\times$3$\times$3 cell cloud centered on the star particle's position.  The reason for this larger cloud is that momentum is a vector quantity and therefore momenta with opposite signs should not contribute to the same cell.  The momentum vector direction for each virtual cloud cell is found by taking the vector from the center of the region to the center of that virtual cell.   Since the virtual cloud is of size $3^3$, it will, in general, overlap $4^3$ cells in the underlying grid.  Figure~\ref{fig:cic} demonstrates graphically how the cloud overlaps with the grid mesh for a two-dimensional example.  The contribution of mass, thermal energy and metals is assumed to be uniformly distributed throughout the $3^3$ cloud, and the addition of these components is very similar to an expanded CIC method.  This method assumes that the underlying mesh overlapped by the cloud has a uniform resolution, i.e. the feedback stencil is never applied over refinement level boundaries, and it assumes that the resolution of the cloud is the same as the underlying mesh.


The magnitude of each cell's momentum vector is found by enforcing the requirement that the total amount of kinetic energy added to the grid follows Equation~(\ref{eq:kinetic_energy}).  Since this depends on the amount of momentum pre-existing on the grid, this step is non-trivial and we describe it in detail below.  We note that the central cell contributes no momentum.  The momentum ($\mathbf{p}_i$), mass ($m_i$) and thermal energy ($e_i$) of each affected grid cell $i$ is updated in this way:

\begin{equation}
\mathbf{p}^a_i = \mathbf{p}^b_i+ \sum\limits_{j}^{ } \Delta p  f_{i,j}\mathbf{x}_j
\end{equation}

\begin{equation}
m^a_i = m^b_i + \sum\limits_{j}^{ }\Delta m f_{i,j}
\end{equation}

\begin{equation}
e^a_i = e^b_i + \sum\limits_{j}^{ } \Delta e f_{i,j},
\end{equation}

\noindent The amount of each gas quantity added to grid cell $i$ involves a sum over all of the virtual `cloud' cells of the feedback zone (denoted with a subscript $j$) that overlap with grid cell $i$.  The quantity $f_{i,j}$ is the fractional volume of that overlap (normalized such that $\sum_{j}^{} f_{i,j} = 1$).
The quantity $\Delta m = m_{\rm inj}/3^3$ is the mass added in each virtual cloud cell $j$, where $m_{\rm inj}$ is the total mass added and $3^3$ is the number of virtual cloud cells.  The quantity $\Delta e$ is the thermal energy added to each virtual cloud cell, which is equal to the total amount of thermal energy injected in the feedback event divided by the number of virtual cloud cells such that $\Delta e = (1 - f_{\rm{kin}}) \frac{E_{\rm{SN}}}{3^3} $.  Finally, the unit directional momentum vector of each cloud cell is given by $\mathbf{x}_j$.  

We now describe how to determine the magnitude of the additional momentum $\Delta p$ for each of the $3^3 - 1 = 28$ virtual cells (the central cell has zero momentum).  The magnitude of each of the vectors must be the same to keep the total net momentum zero.  
The total kinetic energy of the gas surrounding the star particle is a sum over all the affected cells (cells in the grid are denoted with a subscript $i$):
\begin{equation}
\label{eq:thermal_energy_after}
E^a_{\rm{kin}} =   \sum\limits_{i}^{ }  \frac{1}{2}
   \frac{ (\mathbf{p}^b_i+ \sum\limits_{j}^{ } \Delta p  f_{i,j}\mathbf{x_j} )^2}{m^b_i + \sum\limits_{j}^{ }\Delta m f_{i,j}},
\end{equation}

\noindent Because $\Delta p$ is a constant and can be pulled out of the above sum, equation (\ref{eq:thermal_energy_after}) combined with Equation~(\ref{eq:kinetic_energy}) is a quadratic for $\Delta p$, which we can therefore solve:
\begin{equation}
\label{eq:deltap}
\Delta p = \frac{- b + \sqrt{b^2 - 4ac}}{2c},
\end{equation}
where
\begin{equation}
\label{eq:asum}
\begin{split}
a 	&= \sum\limits_{i}^{ } \frac{1}{2} \frac{(\mathbf{p}^b_i)^2}{m^b_i + \sum\limits_{j}^{ }\Delta m f_{i,j}} - E^a_{\rm{kin}}  \\
   	&= \sum\limits_{i}^{ } \frac{1}{2} \frac{(\mathbf{p}^b_i)^2}{m^b_i + \sum\limits_{j}^{ }\Delta m f_{i,j}} - E^b_{\rm{kin}} - f_{\rm{kin}}E_{\rm{SN}}
\end{split}
\end{equation}
\begin{equation}
\label{eq:bsum}
b = \sum\limits_{i}^{ } \frac{\mathbf{p}^b_i \cdot   \sum\limits_{j}^{ } \mathbf{x}_j f_{i,j}}{m^b_i + \sum\limits_{j}^{ }\Delta m f_{i,j}}
\end{equation}
\begin{equation}
\label{eq:csum}
c= \sum\limits_{i}^{ } \frac{1}{2} \frac{(\sum\limits_{j}^{ } \mathbf{x}_j f_{i,j})^2}{m^b_i + \sum\limits_{j}^{ }\Delta m f_{i,j}},
\end{equation}
and the total kinetic energy of the gas in the supernova affected region before the blast is given by:
\begin{equation}
\label{eq:thermal_energy_before}
E^b_{\rm{kin}} = \sum\limits_{i}^{ } \frac{1}{2} \frac{(\mathbf{p}^b_i)^2}{m^b_i}.
\end{equation}

Equations (\ref{eq:deltap}) through (\ref{eq:thermal_energy_before}) provide a complete solution for $\Delta p$.  The quadratic, of course, has a second solution, however, the physical context of the problem requires $\Delta p$ to be a positive value and that in turn requires the solution we have chosen since in general $a \le 0$.  Note that, given the amount of energy and mass to be injected, the scheme has a single free parameter: $f_{\rm kin}$.

\section{Idealized tests}

\label{sec:idealized_tests}

\begin{figure*}
\centering
\includegraphics[scale=0.48] {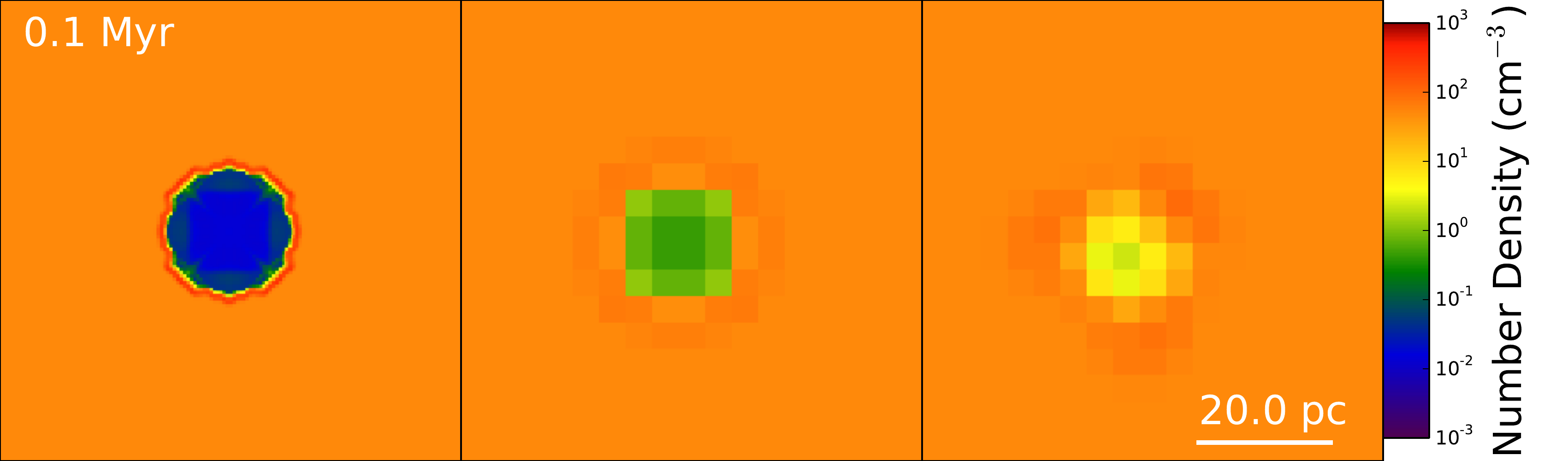}
\caption{Slices of gas density after 0.1 Myr in three different test simulations.  The slices shown pass through the center of the feedback region.  The background density for all three tests is 49 cm$^{-3}$ and the kinetic injection fraction is 0.3.  The left most image shows a test done with the PPM hydrodynamical solver at $\Delta x = 0.5$ pc resolution.  The middle image shows a test also done with PPM, but with $\Delta x = 4.0$ pc resolution.  The right most image shows a test done with the Zeus hydrodynamical solver at $\Delta x = 4.0$ pc resolution.}
\label{fig:slices}
\end{figure*}

To better understand the effect of the kinetic feedback model in galaxy simulations such as the ones that will be discussed in Section \ref{sec:galaxy_simulation} and also to test the implementation, we have conducted a set of idealized test simulations in gas with uniform density.  As we have discussed, the new method introduces a tunable parameter $f_{\rm{kin}}$, the fraction of injected kinetic energy.  Our goal is to understand the behavior of the model given a choice of $f_{\rm{kin}}$ over a wide range conditions typically found in the galaxy simulations for which it is intended.

\subsection{Testing Setup}

\label{sec:testing_setup}

Idealized tests of the new model were done by placing a single star particle within a box with uniform gas density and thermal energy, and no initial bulk motion.  The star particle injects thermal energy, kinetic energy, mass and metals within a single time step into a volume equivalent in size to a $3^3$ cell cloud centered around the star particle as described in Section \ref{sec:kin_algorithm}.  The star particle is chosen to have a mass of 100 \Msun\ and to inject $10^{51}$ ergs of total energy into the surrounding gas.  The amount of mass released is fixed to be 25 \Msun\ and the metal yield is chosen to be 2\% of the star particle's mass.  The values are consistent with the cosmological simulations described in the next section.

The initial gas temperature is set to 100 K.  The particle is placed offset from the box center by half a cell width.  These test simulations were conducted on a uniform, static mesh with a constant cell width in the range of 0.5 pc to 8 pc.  We test background densities between 0.5 and 500 \cm.  

The test simulations include non-equilibrium cooling from six atomic species of hydrogen and helium \citep{abel97, anninos97} in addition to equilibrium metal line cooling dependent on gas density, temperature, electron density and metallicity as interpolated from tables generated by CLOUDY \citep{smith08,ferland98}.  The background medium is initially neutral and composed of 75\% hydrogen and 25\% helium by mass.  The background gas metallicity is set to be 0.1 \Zsun\ in most test simulations, but we will explore the effect of a different background gas metallicity in some tests. 

Runs are conducted with both the PPM and the ZEUS hydrodynamical solvers.  PPM is a piecewise-parabolic method for solving Euler's equations \citep{colella84} and ZEUS is a simple second-order finite difference method \citep{vanleer77,stonenorman92}.  ZEUS has the advantage of being more computationally robust than PPM, but it is less accurate and not energy conserving.  To improve stability in simulations conducted with PPM, we employ the dual energy formalism to maintain consistency between kinetic energy advected by the energy conservation equation and the momentum conservation equation \citep{bryan95}.

The width of the simulation box is fixed to be three times the analytic Sedov-Taylor blast radius, $R_{sedov}$, at the final time of the simulation (typically 1 Myr) with a minimum box width of 50 cells \citep{sedov59,taylor50,draine11}.  For a background density of 1 cm$^{-3}$, a supernova energy of $10^{51}$ ergs and final time of 1 Myr, this gives a box size of 237 pc.  

For these tests, we choose to inject all the energy in a single time step.  Enzo computes time steps from the Courant condition based on the state of the gas at the start of the time step.  This estimation for our idealized tests, where there are no initial gas velocities, will give an initial time step much too long to account for the large velocities that are induced in the first time step by the injection of feedback energy.  An estimate of a shorter time step is necessary to ensure that the bulk flow of gas across the grid in the first time step cannot cross more than one cell width.  The initial timestep, in which all energy is injected, is fixed to be
\begin{equation}
\label{eq:idt}
dt_0 = 0.1 \times \frac{\Delta x}{c_s + v_s},
\end{equation}
\noindent where $\Delta x$ is the diameter of a single grid cell, $c_s$ is the estimated sound speed of gas in the post feedback injection region and $v_s$ is the estimated initial speed of the shock:
\begin{equation}
\label{eq:cs}
c_s = \left((\gamma - 1) \gamma (1-f_{\rm{kin}}) \frac{e_{\rm{SN}}}{\rho \Delta x^3}\right)^{1/2},
\end{equation}

\begin{equation}
\label{eq:vs}
v_s = \left(\frac{2 f_{\rm{kin}} e_{\rm{SN}}}{\rho \Delta x^3}\right)^{1/2},
\end{equation}

\noindent and

\begin{equation}
\label{eq:ms}
\rho = \rho_o + \frac{f_m M_*}{27 \Delta x^3},
\end{equation}

\noindent where $\gamma = 5/3$; $\rho_o$ is initial background density; and $f_m$ is the fraction of star particle mass that is removed from the particle and deposited in the mesh (i.e. $m_{inj} = f_m M_*$).

\subsection{High-resolution reference models}
\label{sec:high_res_ref}   

\begin{figure*}
\centering 
\includegraphics[scale=0.85] {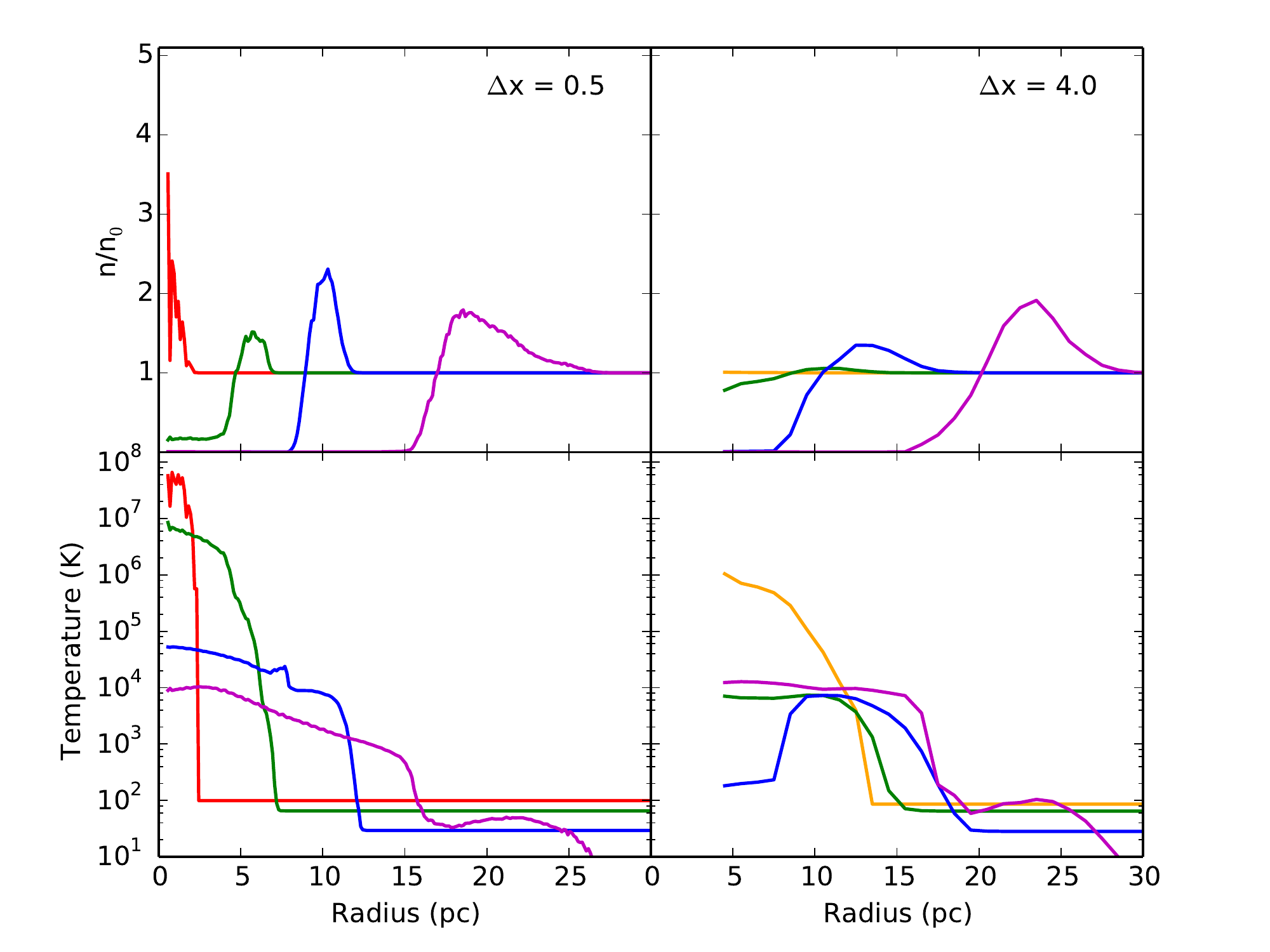}
\caption{Spherically-averaged radial profiles of gas density and density weighted gas temperature for two test simulations of the kinetic feedback model at a high resolution ($\Delta x=0.5$ pc, left) and a low resolution ($\Delta x=4.0$ pc, right).  The gas density is normalized by the background gas density, which is 49 \cm\ in both simulations.  Lines are color coded by time: the time immediately following the first timestep, which is $2.3 \times 10^{-5}$ Myr (red) for the high resolution test and $1.7 \times 10^{-3}$ Myr (orange) for the low resolution test; 0.01 Myr (green); 0.1 Myr (blue); and 1 Myr (magenta).   Both simulations have a kinetic injection fraction of 0.3.}
\label{fig:profiles}
\end{figure*}

\begin{figure*}
\centering
\includegraphics[scale=0.6] {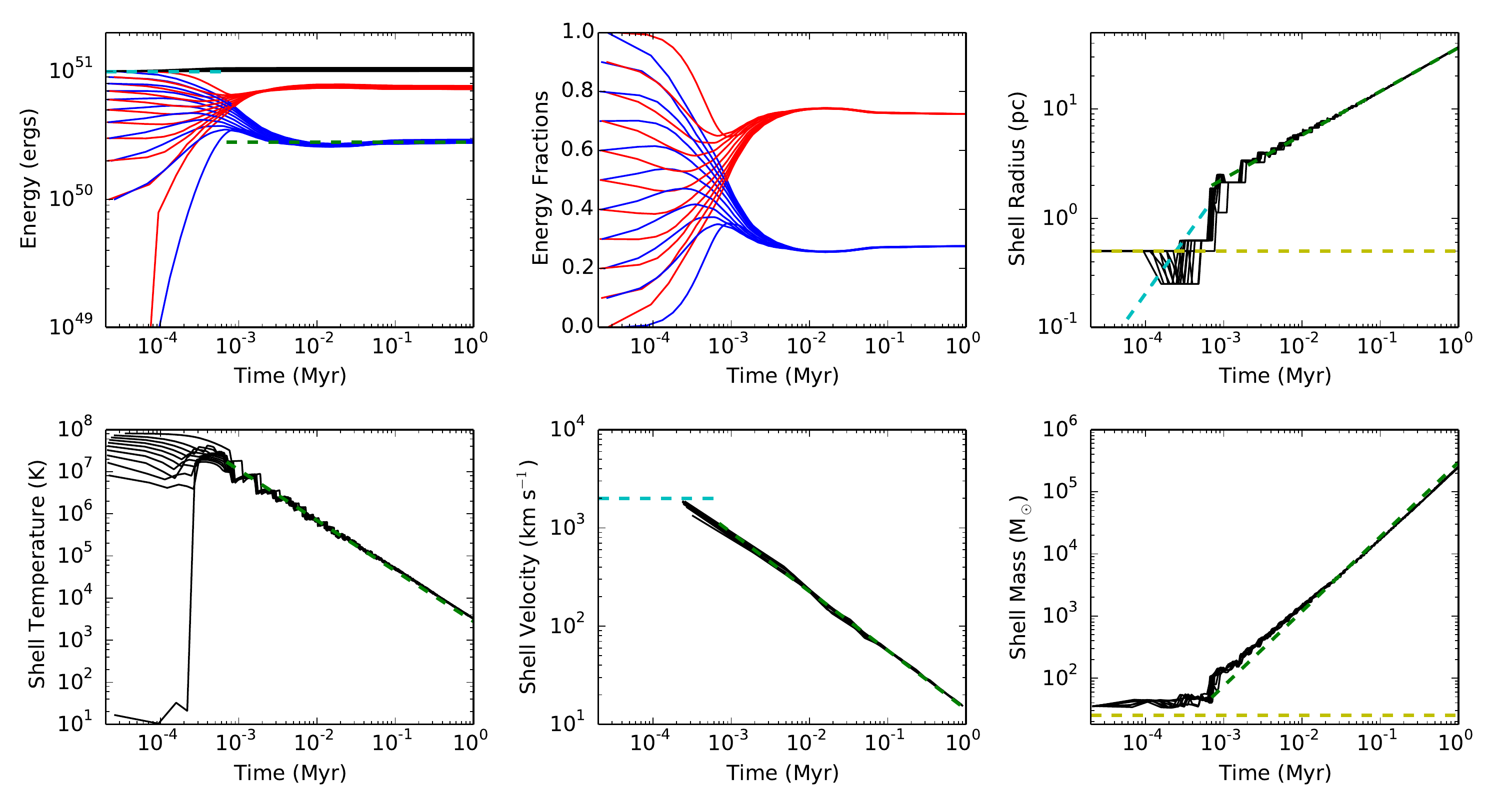}
\caption{High-resolution evolution of quantities derived from adiabatic test simulations of the kinetic feedback model conducted in gas with a background density of 49 cm$^{-3}$.  Eleven simulations with kinetic injection fractions ranging from 0 to 1 are presented.  {\it Top Left:} Excess energy in the simulation box.  Black is the total energy, red is the thermal energy and blue is kinetic energy.  The predicted kinetic energy in the free-expansion phase is shown with a dashed cyan line and for the Sedov-Taylor phase with a dashed green line.  {\it Top Middle:} Fractions of thermal energy (red) and kinetic energy (blue).  {\it Top Right:}  Shell radius.  The dashed yellow line is one cell width and the dashed cyan line is the analytically predicted shock radius during the free-expansion phase.  The dashed green line is the analytically predicted shock radius during the Sedov-Taylor phase.  {\it Bottom Left:} Shell temperature.  The dashed green line is the analytically predicted shock temperature during the Sedov-Taylor phase.  {\it Bottom Middle:}  Shell velocity as calculated from the shell radius presented in the top right panel.  Again, the cyan and green lines are the analytic predictions for this quantity for the free-expansion and Sedov-Taylor phases.  {\it Bottom Right:} Mass in the expanding dense shell.  The dashed yellow line is the total mass the star particle ejects (25 \Msun).  The dashed green line is the analytic prediction of the Sedov solution.}
\label{fig:idealized_nocool}
\end{figure*}

We first explore how the model behaves at high resolution, when all length scales are resolved.  The high resolution tests presented here are done with a fixed cell width of $\Delta x$ = 0.5 pc, which is sufficient to resolve the Sedov-Taylor and (in tests with cooling) the snow-plow phases of supernova remnant evolution with the chosen density.  These high resolution tests are conducted with PPM, which is the more accurate of the two hydrodynamic solvers we consider in this study.  These models will provide context for understanding test models at coarser resolution.

Figures \ref{fig:slices} and \ref{fig:profiles} show the behavior of a typical high-resolution test in the presence of cooling.  The model produces a shock that propagates outward from the star particle.  This shock sweeps up material into a dense shell.  The spherically averaged profile shows the shell over density with respect to the background to be approximately 2 after 1 Myr.  This is less than the overdensity of 4 predicted for a strong shock, in part because of the spherical averaging, but also because the resolution is only marginal to resolve the very narrow shock feature \cite{tasker08}.  As we will see, the total mass in the swept-up shell matches analytic theory.

In high-resolution tests with no cooling, a well-behaved, spherical shock is produced.  In tests with cooling, the shock produced has some degree of asymmetry, the character of which is dependent on the kinetic injection fraction.  In tests with low kinetic injection fractions the asymmetry is dominated by (physical) shell instabilities.  In tests with high kinetic injection fractions ($> 40\%$), it is dominated by the discretization of the kinetic energy injection.  As we will discuss, kinetic fractions of interest for our applications are those less than 30\%, but this limitation should be kept in mind for other applications.

We track the evolution of several quantities in each test simulation.  The shell position is defined as being the radius at which the radial density profile peaks (note that the true shock radius should be slightly greater than this quantity, but this is a good proxy).  The shell temperature is defined as the spherical density-weighted average temperature of the gas at the shell radius, also shown in Figure \ref{fig:profiles}.  The total energy excess in each test simulation is defined as the total energy of the box, both thermal and kinetic, less the energy of the unperturbed background medium, which is entirely thermal.  We define the thermal and kinetic energy excesses in the same way.  The shell mass is defined to be the total mass of gas within the supernova remnant that has an over density relative to the background density greater that 1.1.  These quantities are calculated at each time step utilizing the simulation analysis code yt \citep{turk11}, which was run inline with each Enzo test simulation.  Figure \ref{fig:idealized_nocool} shows the evolution of these quantities for an adiabatic case.

We note that a small amount of excess energy is artificially introduced into these runs by the dual energy formalism used with PPM.  This numerical error is small, on the order of a few percent, and more pronounced in simulations with larger initial kinetic energy fractions.  The origin of this energy is a small over-correction for the negative thermal energies created by numerical errors in the advection of momentum in the first time step.

In the absence of cooling, our test simulations reproduce the analytic Sedov-Taylor solution for supernova remnants almost exactly, where the radius of the shell expands as $t^{2/5}$, regardless of the initial fraction of kinetic energy.  The main deviations from the Sedov-Taylor solution can be seen early in the simulations and appear to roughly approximate the evolution during the free expansion phase that precedes the Sedov-Taylor phase.  This transition is predicted to occur when the mass of the ISM swept-up by the expanding shock becomes equal to the mass ejected by the star.  In our simulations, once the shell mass reaches roughly 50 \Msun, which is twice the ejection mass, there is an abrupt jump in the shell mass above 100 \Msun, and all quantities appear to begin to asymptote to the analytic Sedov-Taylor solution.  The final fraction of kinetic energy found is 25\%, close to the value of 28\% found by integrating the Sedov-Taylor prediction for the radial profile of energy and close to the values of roughly 30\% found in other numerical models of supernova remnants \citep{chevalier74,cox72}.

\begin{figure*}
\centering
\includegraphics[scale=0.55] {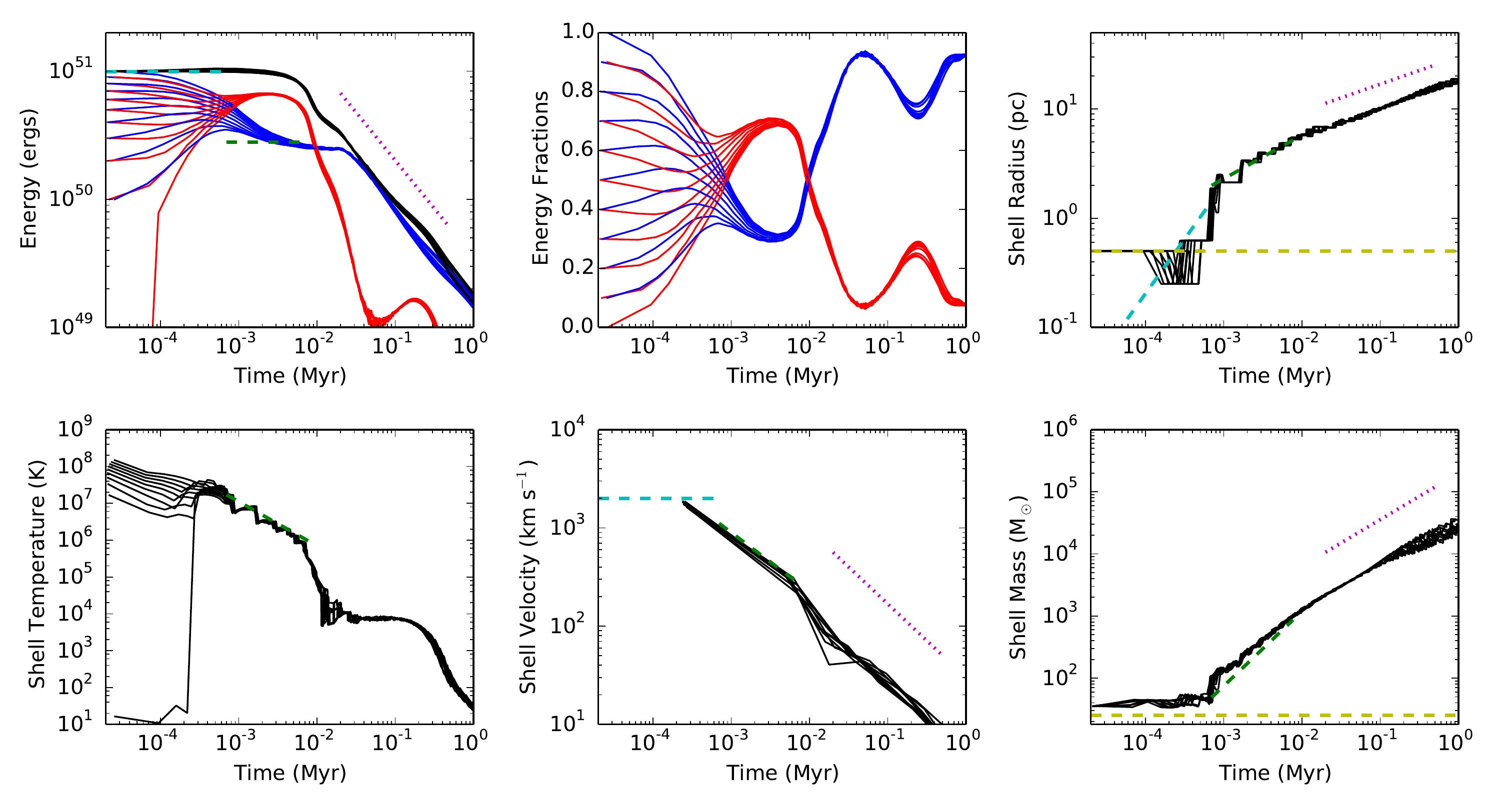}
\caption{High-resolution evolution of quantities derived from test simulations of the kinetic feedback model with cooling.  The quantities presented are the same as those shown in Figure \ref{fig:idealized_nocool}.  The predicted power-law slopes in the momentum-conserving snowplow phase are shown with magenta dotted lines (offset for readability) for the total kinetic energy, shell radius, shell velocity and shell mass.}
\label{fig:idealized_wcool}
\end{figure*}

Figure \ref{fig:idealized_wcool} presents the evolution of the kinetic feedback model in the presence of cooling.  Simulated at high resolution, we see that it takes less than $10^4$ years before cooling causes deviations from the analytic Sedov-Taylor solution and the remnant enters the snow-plow phase.  The shell radius is approximately 6 pc when this happens.  As cooling becomes more and more effective, the temperature of the shell declines sharply and reaches a plateau of $10^4$ K at 0.01 Myr that lasts until 0.2 Myr.  Throughout this phase, the energy drops precipitously due to the rapid cooling of thermal energy.  Therefore, after 0.01 Myr most of the energy remaining in the volume is in kinetic form.  At this point, the decline of the kinetic energy has a power-law slope consistent with the analytically derived value of -3/4 for the momentum-conserving snowplow phase.  Note that, prior to the momentum-conserving snowplow phase, the remnant should go through an adiabatic pressure-driven snowplow phase that has a very similar power-law slope of -4/7; however, in practice that is difficult to distinguish from the slope of the momentum-conserving snowplow.  We run these test for 1 Myr at which time the shell has a radius of almost 20 pc.

The physical relevance of the late time evolution of these tests shell should be considered in the context of a more realistic external medium.  The thin shell is moving at velocities less than the turbulent velocities found in more realistic, inhomogeneous media that are typically of the order of $\sim 10$ km s$^{-1}$ .  The goal of these high-resolution tests, however, is simply to provide context for the resolution study that will be presented in Section \ref{sec:res_study}.

In these high resolution tests, the evolution of gross properties of the supernova remnant is entirely insensitive to the initial fraction of kinetic energy.  In cases both with cooling and without cooling, it takes less than a thousand years for simulations with different initial kinetic fractions to equilibrate to the Sedov-Taylor solution.  After this equilibration period, all simulations behave very similarly, with the exception of some very small differences seen at late time in models with cooling, but these differences are seen well after the simulations have ceased to be physically relevant due to our assumption of a uniform background medium.  The next section will explore how this behavior changes at coarser resolutions, with different background densities and with different numerical solvers.

\subsection{The dual effects of background density and resolution}
 
 \label{sec:res_study}
 
 We have explored the behavior of our model at a relatively high resolution and in high density gas, similar to the density threshold for star formation in our galaxy models.  Now we will examine the effect of changing background density and resolution.  
 
\subsubsection{Energy dissipation}
 
\begin{turnpage}

\begin{figure*}

\includegraphics[scale=0.63] {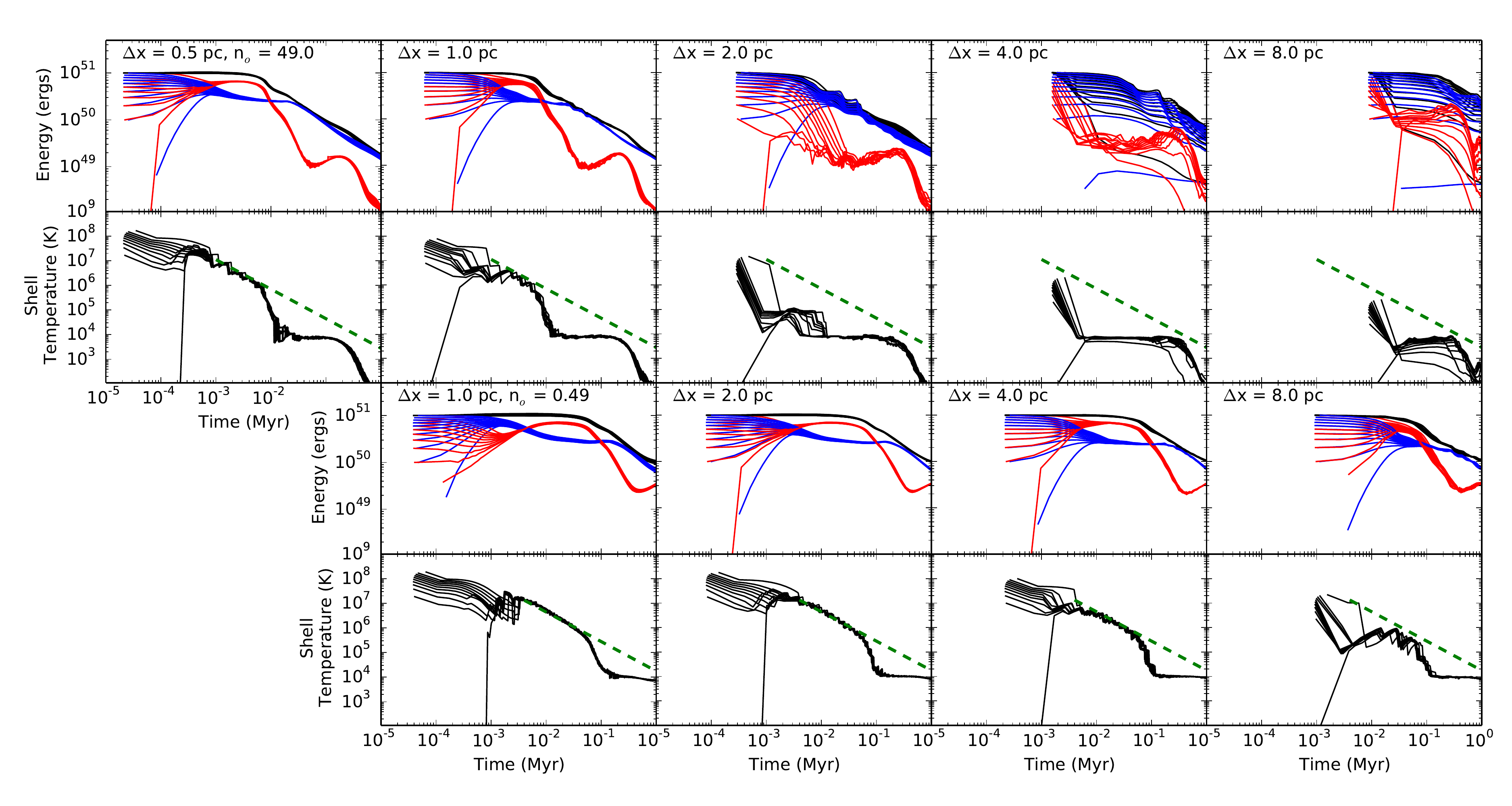}
\caption{Evolution of test simulations at several different resolutions.  Tests are shown at two different background densities: 49 \cm\ (top two rows) and 0.49 \cm\ (bottom two rows).  The quantities shown are the excess energy and shell temperature.  Line colors are the same as in Figure \ref{fig:idealized_nocool}.  Each column presents simulations with at a different resolution $\Delta x$.  Eleven simulations are shown at each density and resolution combination with kinetic injection fractions ranging from one to zero.  Note that the starting times of lower resolution tests are shifted to later times because those tests have longer time steps.}
\label{fig:multires_multiden}

\end{figure*}

\end{turnpage}

Figure \ref{fig:multires_multiden} presents the evolution of our idealized feedback simulations with two different background densities (0.49 and 49 \cm) at a range of spatial resolutions.  For each background density and resolution, we have tested eleven different kinetic injection fractions, varying uniformly from 0.0 to 1.0.
  
First, it is apparent that at the lowest background density, these test simulations behave very similarly, regardless of the kinetic injection fraction or resolution.   All of the low density tests, even the most coarsely resolved, with a cell width of 8 pc, reach a rough approximation to the Sedov-Taylor phase, as indicated by the evolution of the shell temperature.  This brings their later evolution and final state in line with the highest resolution simulations at this density.
 
At higher densities, this uniform behavior between simulations of different kinetic injection fraction and resolution breaks down.  As we have discussed, at very high resolution ($\Delta x=0.5$ pc) for our adopted range of densities, the kinetic injection fraction makes little difference to late time evolution.  This largely holds for simulations with the next two finer resolutions ($\Delta x=1.0$ pc and 2.0 pc).  However, in the two most coarsely resolved sets of simulations where $\Delta x = 4$ pc and 8 pc, simulations with different kinetic injection fractions end with very different total energies.

How can we understand these transitions in behavior?  Simulations that converge to similar behavior regardless of kinetic injection fraction all pass through a well-defined Sedov-Taylor phase, which, in turn, is primarily determined by the temperature to which cells are heated, as high temperature gas ($> 10^6$ K) has a long cooling time.  This temperature is determined by the specific heating rate of the gas in the injection time step, which depends on two factors: (i) the mass of the cell being heated and (ii), the amount of thermal energy injected.  At a fixed density, the cell mass depends on resolution and the amount of thermal energy injected depends on $f_{\rm{kin}}$.  Therefore, in low-density media, cells can easily reach high temperatures because they have small masses and the specific heating rate is high; at high-densities, however, only tests with resolutions $\lesssim$ 2 pc per cell have sufficiently small cell masses to reach this state.  In the high-density case, an increased temperature jump can be achieved by injecting a larger fraction of thermal energy, but this effect appears to be small (except when considering the purely kinetic case) and cannot compensate for the lower resolution. 

A somewhat distinct second effect also contributes to attaining the Sedov-Taylor phase.  The amount of energy that is injected in kinetic form places an upper limit on the total amount of energy that can be lost in the first time step since only thermal energy can be radiated away (directly).  In our model, where the total energy budget is fixed, this goes hand in hand with the long cooling time of high-temperature gas, since the larger the amount of energy locked up in kinetic form, the smaller the increase in the gas temperature.

\subsubsection{Evolutionary timescales}
\label{sec:evo_timescale}

\begin{figure*}
\centering
\includegraphics[scale=0.9] {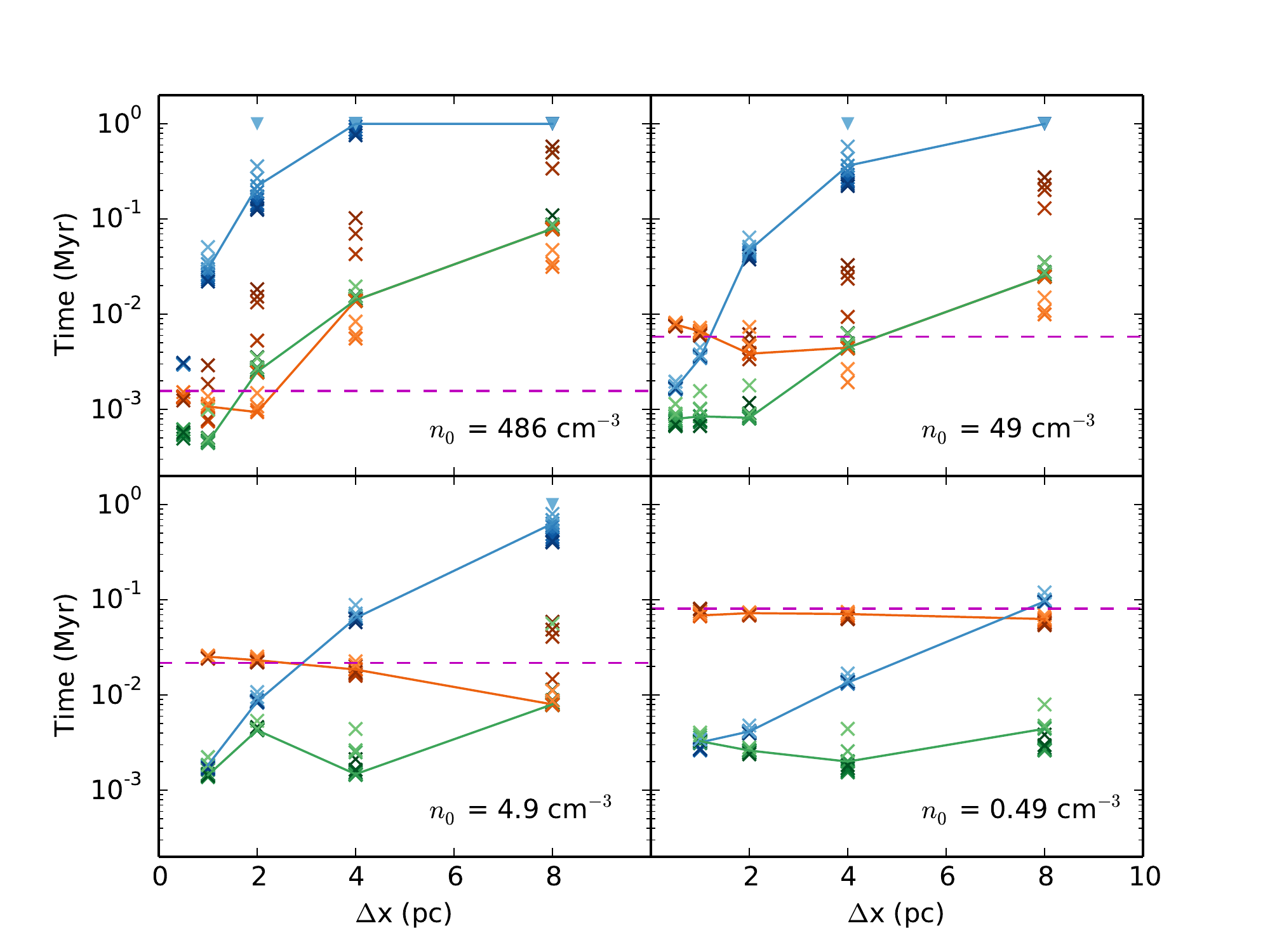}
\caption{The three evolutionary timescales $t_{\rm{\Delta x}}$ (blue), $t_{\rm{eq}}$ (green) and $t_{\rm{snow}}$ (red) for simulations run with varying kinetic injection fractions, $f_{\rm{kin}}$, at four different background densities $n_0$.  Data points with darker shades have larger $f_{\rm{kin}}$.  In some cases, $t_{\rm{eq}}$ exceeds the maximum time of the simulation; these cases are denoted by triangles plotted at 1 Myr.  To guide the eye, lines of the corresponding color connect simulations with $f_{\rm{kin}} = 0.3$ between different resolutions.  An estimate for the transition time between the Sedov-Taylor and pressure-driven snowplow phases from equation \ref{eq:tpds} is shown with a dashed magenta line in each panel.}
\label{fig:time_scales_PPM}
\end{figure*}

 \begin{figure*}
\centering
\includegraphics[scale=0.9] {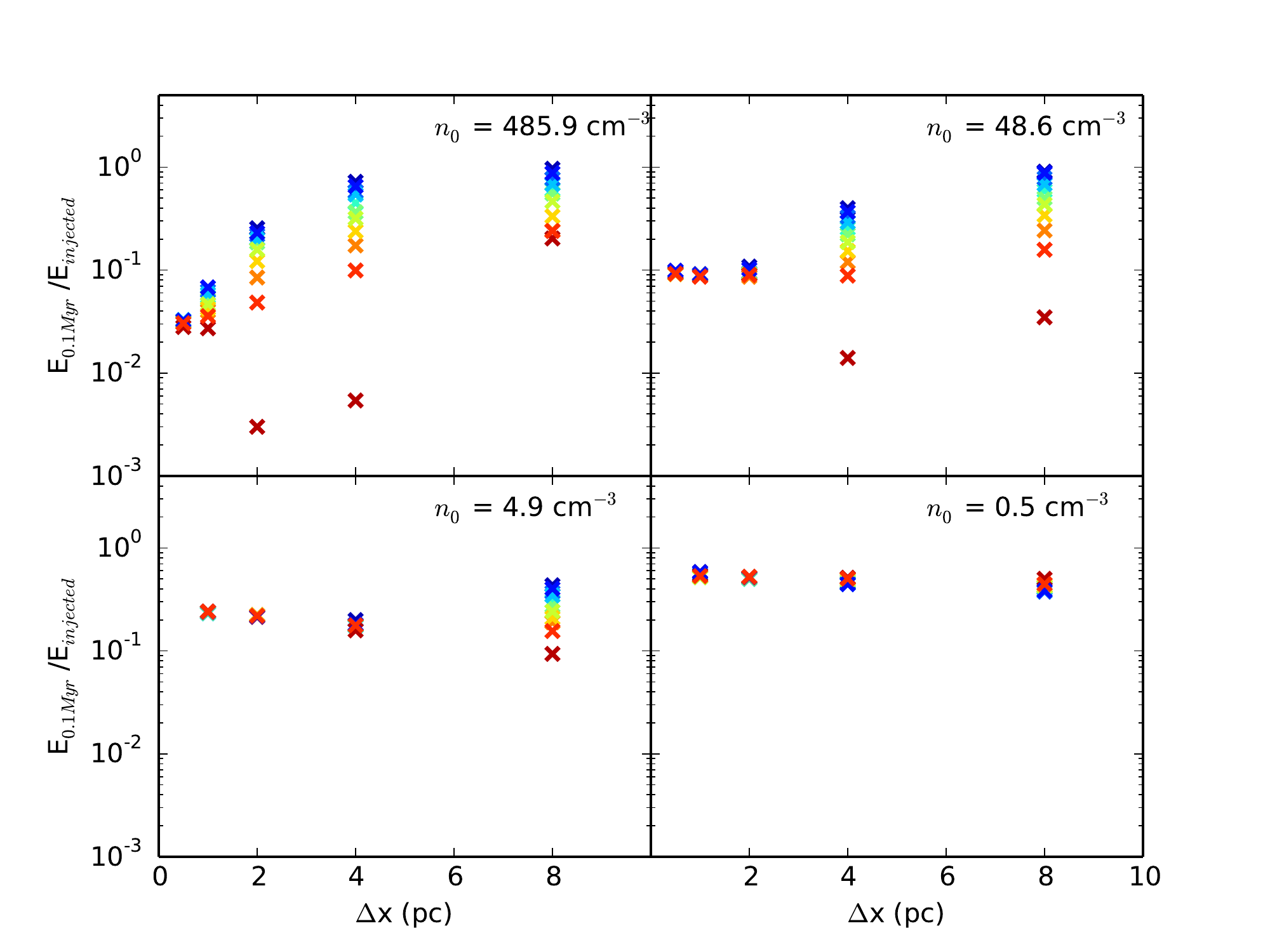}
\caption{The fraction of supernova energy retained in the ISM after 0.1 Myr versus resolution in test simulations run with background densities ranging from 0.5 cm$^{-3}$ to 500 cm$^{-3}$.  Symbols are color coded by the test simulation's initial kinetic injection fraction: dark blue symbols have kinetic injection fractions of one and dark red symbols have fractions of zero. }
\label{fig:final_efrac}
\end{figure*}

To better understand these effects, Figure~\ref{fig:time_scales_PPM} presents characteristic timescales in our test simulations that demonstrate how resolution and radiative cooling interplay with attaining the Sedov-Taylor phase.  Three timescales are considered, related to three evolutionary stages.  The first is a purely numerical stage, during which the blast wave is still localized in the initial feedback injection site, and its evolution is dominated by the discretization of the underlying grid.  Before the heated region can reach a regime dominated by physical processes rather than numerical ones, the remnant must clear this resolution-dominated region.  We estimate the timescale of this phase to be the time it takes the blast radius to reach three times the radius of the energy injection region, or equivalently, the time at which the shell radius reaches $4.5\Delta x$.  This resolution timescale $t_{\rm{\Delta x}}$ can be determined from the evolution of the shell radius, which we track in each test simulation. 
 
The second timescale to consider is the time at which the behavior of the remnant begins to approximate the evolution of the analytic Sedov-Taylor solution.  Physically, the transition from the free-expansion phase to the Sedov-Taylor phase occurs when the mass swept-up by the expanding shock is approximately equal to the mass ejected by the star.  At this point, it is not the initial momentum of the ejecta that is driving expansion, but rather it is the force from the internal thermal pressure of the remnant.  We have chosen to let star particles eject 25 \Msun\ into the ISM, however as we have discussed, the initial evolution is dominated by numerical effects of the local grid.  This local grid contains mass and it would be more appropriate in our context to consider a numerical ejection mass equal to the ejection mass plus the mass of ISM in the initial feedback zone of $3^3$ cells.  We therefore define the Sedov-Taylor equilibrium timescale to be the time $t_{\rm{eq}}$ when the swept-up mass is equal to twice this numerical ejection mass.  We can estimate the mass of swept-up ISM from the shock radius and the background density.  
  
The third relevant timescale is the time at which the remnant starts to cool radiatively, and enters the snow-plow phase.  Following \citet{draine11} we define this timescale ($t_{\rm{snow}}$) to be the time at which the energy of the remnant has dropped by one third of its original value.  Note that this typically happens well before the shock temperature reaches the cooling floor, and therefore it is not dependent on our choice of cooling floor.

  \begin{figure*}
\centering
\includegraphics[scale=0.5] {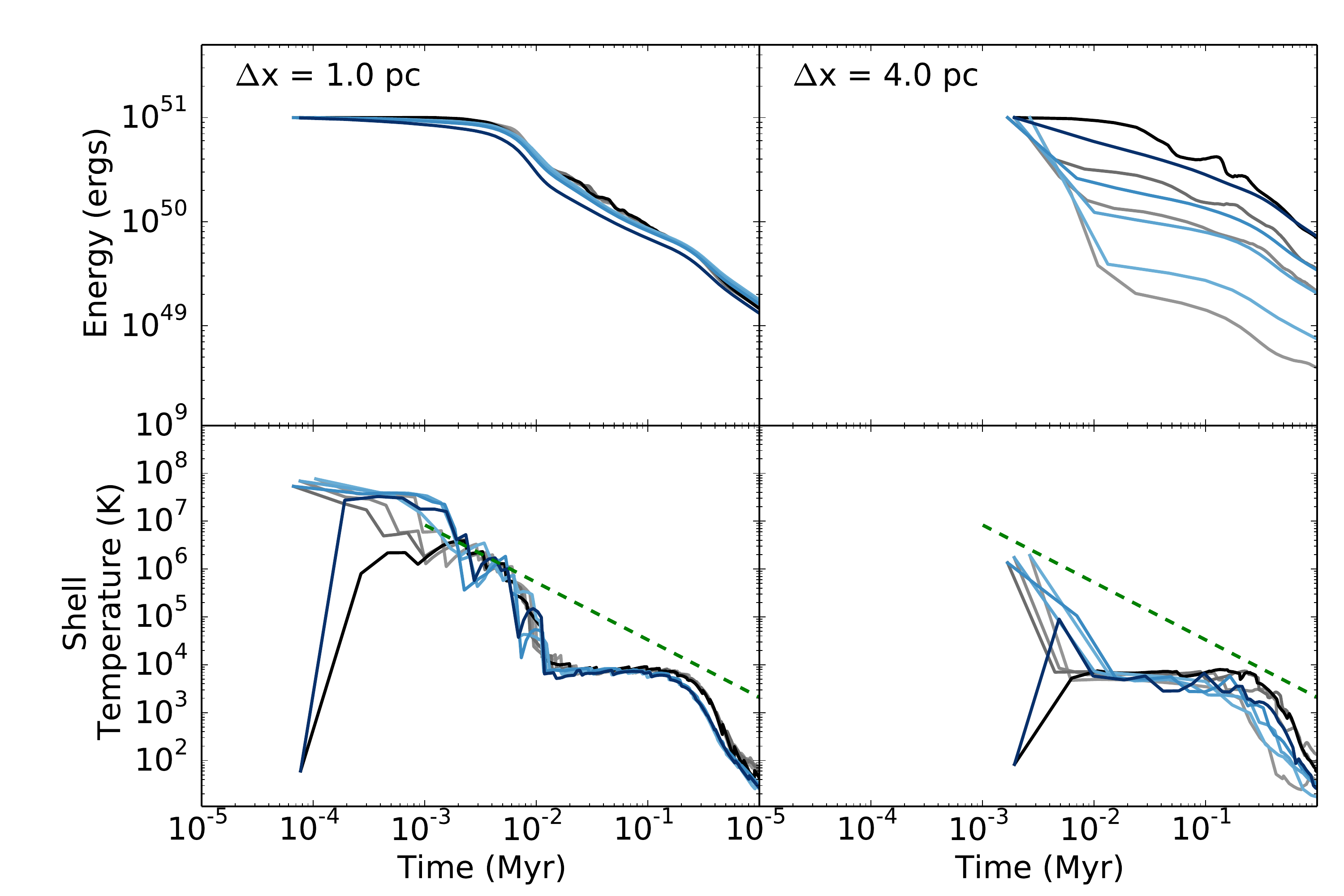}
\caption[]{Comparison between test simulations run with PPM (grey lines) and Zeus (blue lines).  Darker lines have a greater fraction of injected kinetic energy.  The kinetic injection fractions shown are 0.0, 0.1, 0.3 and 1.0.  The top row shows the evolution of the energy excess in the simulation volume and the bottom row shows the shell temperature, just as is shown in Figure \ref{fig:multires_multiden}.  Test simulations are shown at two different resolutions: $\Delta x = 1.0$ pc (left) and $\Delta x = 4.0$ pc (right).  As before, the dashed green line is the predicted Sedov shock-temperature.}
\label{fig:zeus_comp}
\end{figure*}

 From Figure~\ref{fig:time_scales_PPM}, we see that $t_{\rm{\Delta x}}$ rises with increasing cell width $\Delta x$, and the steepness of this rise increases with increasing background density.  Simulations with low kinetic injection fractions also have slightly longer resolution timescales.  These trends can be explained by the expectation that simulations with larger cells and simulations where the supernova ejecta have lower initial velocities will take longer to expand beyond the resolution dominated region.

At the lowest density ($n_0 = 0.49$ cm$^{-3}$), $t_{\rm{eq}}$ and $t_{\rm{snow}}$ are constant over $\Delta x$ and $t_{\rm{snow}}$ remains greater than $t_{\rm{\Delta x}}$ for all except the very largest $\Delta x$.  In this regime, the supernova remnant equilibrates to the Sedov-Taylor solution before energy losses due to cooling become significant.

At the two next higher densities ($n_0 = 4.9$ cm$^{-3}$ and 49.0 cm$^{-3}$), $t_{\rm{\Delta x}}$ becomes significantly greater than $t_{\rm{snow}}$ at coarser resolutions.  This means that the supernova remnant has not yet cleared the resolution dominated region before energy losses due to cooling become significant.  The result is that the snow-plow time becomes much more dependent on the kinetic injection fraction.  Simulations with higher kinetic injection fractions have significantly larger $t_{\rm{snow}}$ when $t_{\rm{\Delta x}}$ is much greater than $t_{\rm{snow}}$.  This is because the conversion of kinetic to thermal energy in the expanding shock is impeded when $t_{\rm{\Delta x}}$ is long, and so with more energy locked up in kinetic form, less thermal energy is available to be dissipated and the time it takes for total energy to drop significantly is much longer.

At the very highest density probed ($n_0 = 490$ cm$^{-3}$), $t_{\rm{\Delta x}}$ is longer than $t_{\rm{snow}}$ at all resolutions tested.  In this regime, the Sedov-Taylor phase and the snow-plow phase cannot be well separated.  The snow-plow timescale and the Sedov-Taylor equilibrium timescale happen at the same time because the conversion between thermal and kinetic energies is so inefficient when $t_{\rm{\Delta x}}$ is long that the remnant cannot equilibrate before cooling begins to take over.  The increase of $t_{\rm{snow}}$ with increasing $\Delta x$ is apparent at the two highest background densities.  From Figure \ref{fig:multires_multiden}, this trend is likely due to the low initial temperatures reached by heated gas.  In these test simulations the specific heating rate is low at low resolution, resulting in lower cooling rates and a longer $t_{\rm{snow}}$. 
 
The result of this blending of numerically dependent and physically dependent evolutionary timescales is that the total amount of energy and momentum retained in the gas becomes dependent on the kinetic injection fraction in a resolution dependent way.  Figure \ref{fig:final_efrac} shows this effect.  In a low density medium, the amount of energy retained in the gas after 0.1 Myr changes little with resolution and kinetic injection fraction.  Moving to higher density media, the amount of energy retained at coarser resolution begins to vary with the kinetic injection fraction, where models with higher kinetic injection fractions lose less energy.  In models with background densities of 49 and 490 \cm, purely thermal models lose an order of magnitude more energy than models with some injection of kinetic energy.  Models that have very high kinetic injection fractions retain too much total energy.

 \subsection{How secondary effects can change energy dissipation}
 
 \label{sec:second_order}
 
  Secondary effects can change these trends by modifying the rate of dissipation of thermal energy.  In tests with purely thermal energy injection, these effects can potentially be quite significant and cause a greater under or overestimate of the retained energy at late times.  As an example, consider the effect of increasing the background metallicity.  At high resolution, test runs with a higher background metallicity and different kinetic injection fractions converge in much the same way as they do for tests at lower background metallicity, although at an overall lower energy and temperature.  In Figure \ref{fig:final_efrac}, we see that the purely thermal case underestimates the amount of energy remaining in the gas at 0.1 Myr by nearly 84\% with 4 pc resolution (where the background density is 50 \cm\ and metallicity is 0.1 \Zsun).  We found that in similar test simulations with a higher background metallicity of 1 \Zsun, the purely thermal case undershoots the total energy at 0.1 Myr by nearly 95\%.

In the case of higher metallicity, the cause of enhanced energy loss has its origin in the physical model employed for the cooling that is then magnified by numerical effects.  It can also be the case that purely numerical effects can produce differences.  Figure \ref{fig:zeus_comp} compares tests run with the PPM hydrodynamical solver, which has been used for all the test simulations up to this point, to tests run with the ZEUS hydrodynamical solver.  The physical model in these two sets of test cases are identical; the differences seen between simulations with the same kinetic injection fraction are due to numerical effects.  We can see that at high resolution the differences between the PPM and ZEUS solvers are small.  At coarse resolution, there is a divergence in behavior with the kinetic injection fraction $f_{\rm{kin}}$.  Tests with some injection of kinetic energy, even as little as 10\%, end up with the nearly the same amount of thermal energy at late times; however, it should be noted that at earlier times there are differences.  The greatest difference, again, can be seen in the purely thermal case, where tests run with both solvers underestimate the higher resolution result, but the PPM model does so by a greater degree.  The differences between these two solvers likely have their origin in the competing effects of imperfect energy conservation and increased diffusivity.

\subsection{A scheme for varying the kinetic fraction with resolution}
\label{sec:kinf_var}

 \begin{figure}
\centering
\includegraphics[scale=0.7] {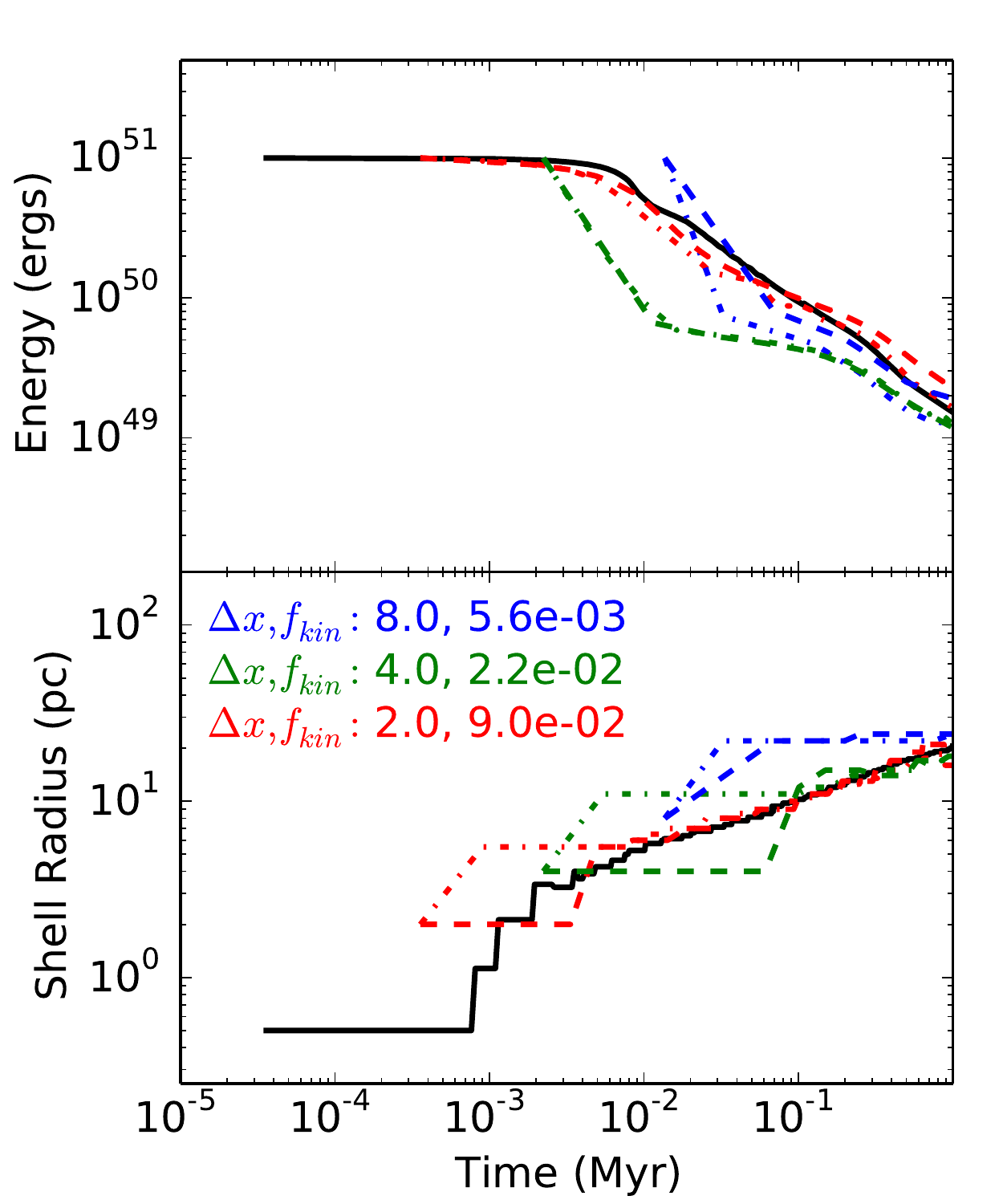}
\caption[]{Test simulations with varying kinetic fractions that are determined from the simulation resolution according to equation \ref{eq:kinf}.  All tests are compared to a high resolution test ($\Delta x = 0.5$ pc) conducted with the PPM solver and $f_{\rm{kin}} = 0$ (black).  Lower resolution tests were conducted with PPM (dash-dot lines) and ZEUS (dashed lines).  The top panel shows the energy excess in the simulation volume and the bottom panel shows the shell radius.  The background density for this test is 49 \cm.  The resolutions tested are indicated in parsecs along with corresponding kinetic injection fractions.}
\label{fig:kinf_var}
\end{figure}

\begin{table*}
\centering
  \caption{Summary of Properties of Galaxy Simulations. \label{tab}}
  \begin{tabular}{@{}lcccccccc}
  \hline
   &$f_{\rm{kin}}$ & $M_{tot}$& $M_*$ & $r_{200}$ &$r_{1/2}$&$M_{1/2}$ &$\langle \rm{Log}(\frac{Z}{Z_\odot}) \rangle$ & $\sigma_Z$\\
   & & ($10^9 M_\odot$) & ($10^5 M_\odot$) &(kpc) &(pc)&($10^7 M_\odot$) &\\
 \hline
 R10 &  - &  1.55 &  14.3  &  23.7 &704 &  3.05& -0.52& 0.86\\
  R10-kinf-0$^*$ &0 & 0.980 & 48.4  &  13.9 &441 &1.88 & 0.10&0.45\\
   R10-kinf-0.01 &0.01 & 1.61& 1.79&24.1  &607 & 2.95& -1.26& 0.43\\
  R10-kinf-0.05 &0.05 & 1.55 & 1.16  &23.7&781&3.57&-1.70 &0.37\\
  R10-kinf-0.1&0.1 & 1.54&1.38& 23.6&572&2.14 &-1.71&0.42\\
  R10-kinf-0.3  &0.3  & 1.61&1.97  &24.0 &1051 &5.28 &-1.68& 0.35\\
  R10-var$^{*,1}$  & - & 1.02	& 3.55 &12.9 & -& -&-0.93&0.51 \\
R10-var-dtlim & - & 1.59& 2.07& 24.0& 717& 3.32& -1.20& 0.47\\
R43-var-dtlim & - &1.56 &3.25 & 23.8&750 & 3.62&-1.19 &0.44 \\
R173-var-dtlim & - & 1.61& 2.38&24.1 & 583& 2.18&-1.53 & 0.41 \\
  \hline

\end{tabular}

 \vspace{5 mm}
\raggedright Note: The quantities presented in each column are (1) the constant kinetic injection fraction (if applicable), (2) the total mass within $r_{200}$, (3) the total stellar mass within $r_{200}$, (4) $r_{200}$, the radius within which the mean halo density is 200 times the critical density of the universe, (5) the radius enclosing half the stellar mass, (6) the total mass within $r_{1/2}$, (7) the mass-weighted mean of the star particle metallicities and (8) the standard deviation of Log(Z/\Zsun).

\vspace{5 mm}
\footnotesize * Simulation R10-kinf-0 was only run to a redshift of 0.92 and R10-var was run to redshift 1.14; therefore, the quantities presented here are for the target halo at this redshift.

\footnotesize 1 Simulation R10-kinf-var is in the process of a luminous merger at its final redshift that makes a meaningful half-light radius difficult to compute.
\end{table*}

We have shown how energetic losses depend on resolution with our implementation of kinetic feedback.  In this section, we present an analytically-motivated, but simulation-verified, scaling for $f_{\rm kin} (\Delta x, n_o)$ as a function of resolution and local density.  Rather than doing this directly from the numerical tests present in Section~\ref{sec:res_study} (because their precision is limited by secondary effects), we develop an analytical model and then demonstrate that it matches our test simulations.

The scaling we adopt is based on a comparison between the shock radius at the transition point to the pressure-driven snowplow phase ($R_{\rm{PDS}}$), and the resolution of the grid ($\Delta x$).  In Section~\ref{sec:evo_timescale}, we defined the timescale $t_{\rm{\Delta x}}$ as the point when the shell radius was $4.5 \Delta x$ and argued that this represented a purely numerical resolution timescale for the system.  The $f_{\rm{kin}}$-dependent furcation of energetic losses occurs when the timescale for significant energetic losses exceeded $t_{\rm{\Delta x}}$.  

We therefore propose the following procedure.  In gas where $R_{\rm{PDS}} > 4.5 \Delta x$, we consider the Sedov-Taylor regime to be well resolved and choose $f_{\rm{kin}} = 0$.  Conversely, in gas where $R_{\rm{PDS}} < 4.5 \Delta x$, we estimate $f_{kin}$ to be the ratio of the kinetic energy in the system at the point in the snowplow phase when the shock radius equals $R_{\rm{\Delta x}}$ to the total amount of energy produced by the system.  We note that in cases where $4.5 \Delta x >> R_{\rm{PDS}}$ radiative cooling will have resulted in the significant loss of thermal energy.  We choose not to account for this loss and instead demonstrate that in the low-resolution regime where this is important, early rapid cooling quickly compensates for this.   The kinetic injection fraction we therefore adopt is

\begin{equation}
\label{eq:kinf}
f_{\rm{kin}} = 3.97 \times 10^{-6} \mu n_o R_{\rm{PDS}}^7 t_{\rm{PDS}}^{-2} \Delta x^{-2} E_{51}^{-1},
\end{equation}

\noindent where $\mu$ is the mean molecular weight of the surrounding gas; $n_o$ is the density of the surrounding medium in units of \cm; $t_{\rm{PDS}}$ is the time of the transition of between the Sedov-Taylor phase and the pressure-driven snowplow phase normalized by $10^3$ years; $R_{\rm{PDS}}$ is the radius of the shock at $t_{\rm{PDS}}$ in pc; $\Delta x$ is the cell width in pc; and $E_{51}$ supernova energy normalized by $10^{51}$ ergs.  

With an assumption for the cooling rate, it is possible to estimate the time and radius at the transition between the energy-conserving Sedov-Taylor phase and the pressure-driven snowplow phase.  Here, we follow \citet{cioffishull91} who assumed a power law cooling function when dominated by metal cooling, and bremsstrahlung cooling for very low metallicity free gas.  \citet{cioffishull91} estimate the transition time between the Sedov-Taylor phase and the pressure-drive snowplow phase as the time of the temperature `sag' as defined by \citet{cox72}.  We adopt a transition time when radiative cooling begins to dominate the dynamics of the system, which happens some time after the temperature `sag' and is also described by \citet{cox72}.  
\citet{thornton98} find a break in the evolution of of supernova remnants with metallicities below 0.01 \Zsun.  Therefore we adopt the result for a metal dominated cooling curve for metallicities above 0.01 \Zsun\ and the result for the bremsstrahlung dominated cooling curve for metallicities below 0.01 \Zsun:

\begin{equation}
\label{eq:tpds}
t_{\rm{PDS}} = \left \{ \begin{array}{rl}
3.06 \times 10^2 E_{51}^{1/8} n_o^{-3/4} &\mbox{ if $Z<0.01$} \\
26.5 E_{51}^{3/14} n_o^{-4/7} Z^{-5/14} &\mbox{ if $Z \ge 0.01$}
\end{array} \right.
\end{equation}

\begin{equation}
\label{eq:Rpds}
R_{\rm{PDS}} = \left \{ \begin{array}{rl}
49.3 E_{51}^{1/4} n_o^{-1/2} &\mbox{ if $Z<0.01$} \\
18.5 E_{51}^{2/7} n_o^{-3/7} Z^{-1/7} &\mbox{ if $Z \ge 0.01$}.
\end{array} \right.
\end{equation}

\noindent In this form, $t_{\rm{PDS}}$ is normalized by $10^3$ years, $R_{\rm{PDS}}$ is in units of parsecs, Z is the gas metallicity in solar units and $n_o$ and $E_{51}$ are defined as before.

Figure \ref{fig:kinf_var} shows an example of this approach applied to the same test problem at a variety of resolutions and demonstrates how it compares to the high-resolution case.  All of the tests presented show cases where we do not consider the Sedov-Taylor phase to be resolved, however, the criterion we adopt for this determination is generous enough that in some cases, significant energetic losses have yet to occur.  In these cases, the high-resolution behavior is, unsurprisingly, reproduced well.  At the other extreme, there are cases where very significant energetic losses have taken place in the high-resolution model by the time the coarse resolution model has begun.  In these cases, the total energy is overestimated in the first time step, however, significant cooling by the second time step quickly corrects this and the overestimate of the shell radius is of order the cell width.  An intermediate case, where the resolution is such that energetic losses are just on the verge of becoming significant provides perhaps the most problematic case, where cooling causes the energy to drop much more quickly than in the high-resolution case, and therefore convergence to the high-resolution case is delayed.  Still, we conclude that our analytic model for $f_{\rm kin}$ provides a reasonable approximation to the high-resolution results.

\section{Example application: Cosmological dwarf galaxy}
\label{sec:galaxy_simulation}

Idealized tests are useful for understanding the essential scaling; however, more realistic tests are necessary to see how the method works in practice.  In this section, we apply our method to a cosmological zoom-in simulation of a low-mass galaxy.  We carry out runs with several different kinetic injection fractions in the range of those that yield reasonable final energy fractions in our idealized tests with similar background densities and cell resolutions.  In addition, we carry out runs with the analytic resolution-dependent scheme for kinetic energy injection that we developed in the previous section.

\subsection{Galaxy simulation initial conditions and physical prescriptions}
\label{sec:galaxy_methods}

We model a low-mass dwarf system in a cosmological zoom-in simulation as described in \citet{simpson13}.  The simulated halo has a mass within $r_{200}$ of $1.55 \times 10^9$ \Msun\ at $z=0$ and is situated in an isolated environment within a cosmological box 4 h$^{-1}$Mpc across \citep[$\Omega_m=0.274$, $\Omega_b=0.0456$, $\Omega_{\Lambda}=0.726$ and $h=0.705$]{hinshaw09}.  
High-resolution dark matter particles have a mass of $5.4 \times 10^3$ \Msun. 
Adaptive mesh refinement (AMR) of the cartesian mesh (on which the hydrodynamical calculation is done) is allowed down a total of 12 levels from the root grid, giving a minimum allowed cell width of 10.8 comoving pc.  
In a subset of simulations, we restrict the maximum number of AMR levels to 10 and 8, correspond to minimum cell widths of 43.2 and 173 pc respectively.  The purpose of these simulations is to explore the effect of limiting the resolution in the hydrodynamic calculation while utilizing the resolution dependent scaling for the kinetic injection fraction described in Section \ref{sec:kinf_var}.
AMR is restricted to cells containing high-resolution dark matter particles that are pre-selected from a dark-matter only simulation of the refined initial conditions and were found to end the simulation within $3r_{200}$ of the final halo.

A range of baryonic physics is included in these simulations: equilibrium metal line cooling \citep{smith08}; non-equilibrium formation and cooling from nine atomic and molecular species \citep{abel97,anninos97}; a metagalactic photoionizing and photodissociating background \citep{haardtmadau01,haardtmadau11}; a simple, single-cell self-shielding prescription \citep{simpson13}, and a star formation algorithm that creates star particles with a maximum mass of 100 \Msun\ based on local cell properties such as gas density, cooling time, dynamical time and divergence \citep{cenostriker92}.  Complete details of these implementations can be found in \citet{simpson13}; the models presented here are identical to the model R10 from that study, except that the single-cell thermal feedback prescription has been replaced with the new prescription described here.  We test both constant kinetic injection fractions (between 0.3 and 0.01, listed in Table \ref{tab}) and a resolution-varying kinetic injection fraction, described in Section \ref{sec:kinf_var}.

The new feedback model allows for the injection of energy on short timescales, which was challenging with the feedback model used by \citet{simpson13} because of the large energy gradient created by the single-cell energy injection.  In the example simulations described here, we choose to inject all the energy, mass and metals produced by supernova feedback in a single time step.  We also choose to inject the supernova energy promptly, immediately after the creation of the star particle, mimicking Type II supernovae (and stellar winds), but neglecting the significant contributions from Type Ia supernovae, which can be delayed from the epoch of their progenitor's creation by as much as a Gyr.  

As in R10, star particles return 25\% of their mass to the ISM during feedback.  They also return metals to the ISM during feedback; the supernova metal yield is assumed to be 2\%.  All metals produced by star particles contribute to a single mean metallicity field.  The total amount of specific supernova energy produced by each star particle is fixed to be the same as in simulation R10.  This value is chosen such that 150 \Msun\ of star particles inject $10^{51}$ ergs of energy into the ISM.  For a well-sampled IMF, the energy (and mass return) from star particles depends on the choice of IMF and assumptions about stellar evolution, however, our star particle mass is too low to be considered a well sampled stellar population; therefore, our choice of energy return is chosen to simply to be consistent with a range of values determined by other works \citep[e.g.][]{jungwiert01,kroupa07,oppenheimer08,conroy10}.

The aspect of the energetic supernova feedback that differs in these new simulations from those presented in \citep{simpson13} is the fraction of energy that takes an initially kinetic rather than thermal form.  This value is a free parameter, $f_{\rm{kin}}$, that we will modify and test.

A few adjustments not detailed in Section \ref{sec:kin_algorithm} were required for the galaxy application of the kinetic feedback model.  Enzo partitions the simulation domain into simply connected regions with uniform resolution that are called `grids.'  Enzo solves Euler's equations on these grids separately as a strategy for distributing the required calculations across many processors working in parallel.  During the evolution of a single time step on the coarsest level of refinement, called the root grid, spatially contiguous grids do not communicate with one another.  Therefore, star particles whose feedback zones straddle grid boundaries cannot properly deposit their energy, mass and metals into the gas.  A solution to this issue is in progress within the Enzo development community, but for the current runs, we have simply shifted the feedback zones away from grid boundaries as needed in order to fit entire feedback zones onto their host grids.  The shift is typically only one cell in a given direction, which is one third the width of the feedback zone.  

Another adjustment is to do with the time step.  In the idealized tests described in Section~\ref{sec:idealized_tests}, the time step of the grid is adjusted in anticipation of the energetic feedback.  In the galaxy simulations, gas will have a non-zero bulk velocity due to the gravitational forces present in the calculation.  However, due to the small gravitational potential of the forming dwarf galaxy, a courant limited time step that accounts for the sound speed and bulk velocity of gas often exceeds $t_{\rm{PDS}}$, the timescale for the transition between the Sedov-Taylor and pressure-driven snowplow phases of supernova evolution (see Equation \ref{eq:tpds}).  Therefore, in our simulation with a resolution dependent kinetic energy injection scheme, we have conducted a test (R10-kinf-var-dtlim) where the time step of grids containing newly-formed star particles are limited to be one-tenth of $t_{\rm{PDS}}$ at the threshold density for star formation in solar metallicity gas.  Since the density threshold for star formation used varies with redshift, this limiting timestep is redshift dependent:

\begin{equation}
\label{eq:dt_lim}
\Delta t_{\rm{lim}} = 6.35 \times 10^3 \times (z+1)^{-12/7} \ \rm{[yr]}.
\end{equation}

We will explore the effect of this time step constraint in Section \ref{sec:dtlim}.  

\subsection{Gas and star formation histories}
\label{sec:sfr}

\begin{figure}
\centering
\includegraphics[scale=0.55] {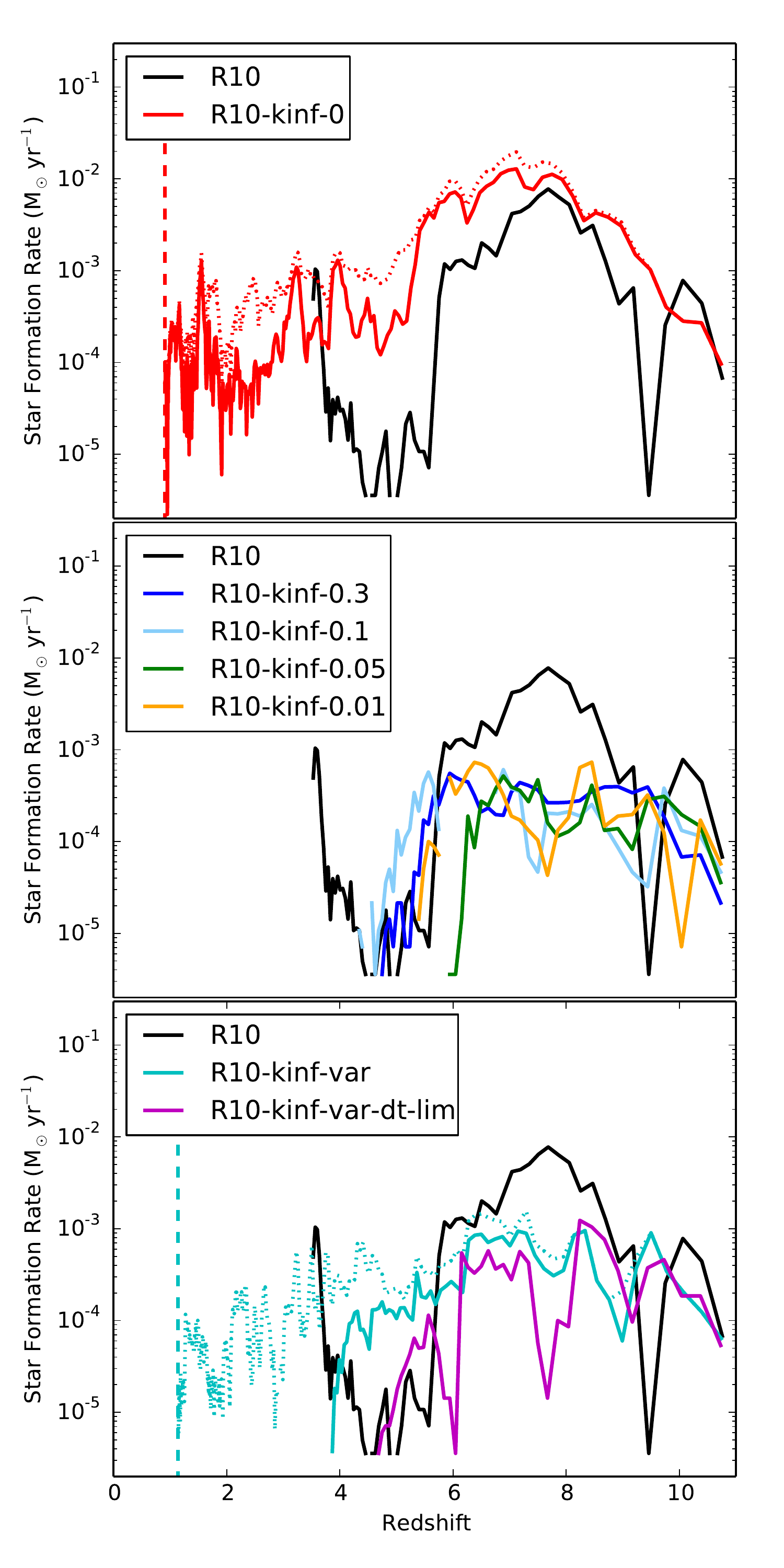}
\caption{Star formation rates in simulations with no kinetic energy injection (top); a constant fraction of kinetic energy injection (middle); and a resolution-dependent kinetic injection fraction (bottom).  Solid lines show the cumulative star formation history of all star particles that end the simulation within $r_{200}$.  Simulations R10-kin-0 (top) and R10-kinf-var (bottom) were stopped close to $z=1$ (see Table \ref{tab}).  Their final redshift is indicated by vertical dashed lines.  For these two simulations, the global star formation rate is shown (dotted lines) in addition to the star formation rate of the final, main halo.}
\label{fig:sfr_comp}
\end{figure}

Star formation in galaxy simulations with the new feedback model exhibits many similarities and some key differences relative to each other and to R10 from \citet{simpson13}.  Table \ref{tab} summarizes some of the properties of the final halos produced in these simulations.  

Figure \ref{fig:sfr_comp} shows the star formation rate in all the galaxy simulations presented here.  All begin by forming stars around $z \sim 11$, when molecular hydrogen cooling in these primordial gas halos becomes effective (see \citet{simpson13} for more details).   The star formation rates in the runs without momentum input (R10 and R10-kinf-0) steadily rise until reionization, when they begin to decline.  Star formation in R10-kinf-0 continues to late times at a reduced rate from its pre-reionization peak.  However, runs with momentum input, both constant and time varying, behave very differently: once the star formation rate rises to a value of approximately $3 \times 10^{-4}$ \Msun yr$^{-1}$ it plateaus and remains constant until reionization.  Post-reionization, the star formation rate quickly drops to zero in almost all cases with (the exception being R10-kinf-var, which we will discuss in Section \ref{sec:dtlim}).

\begin{figure}
\centering
\includegraphics[scale=0.55] {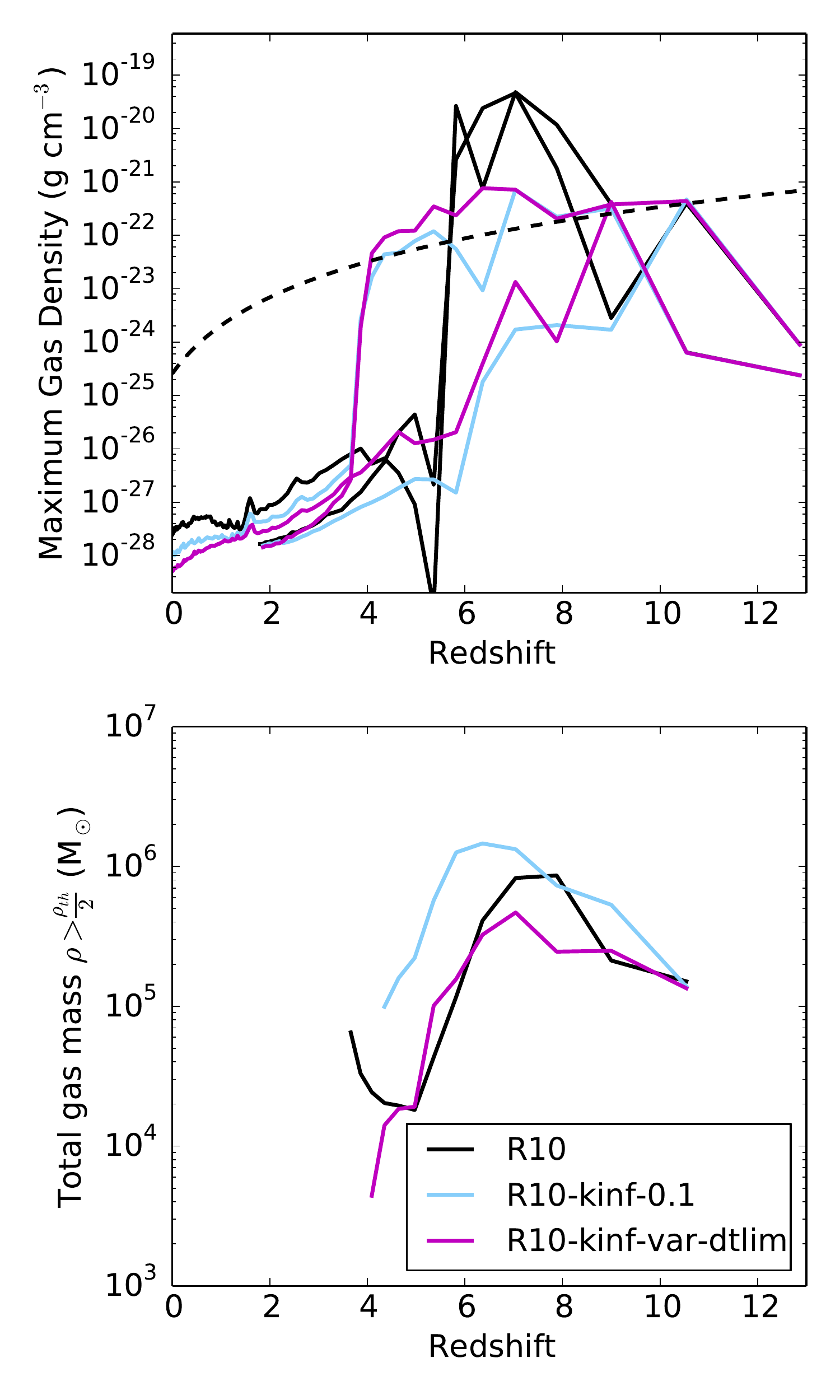}
\caption[]{Evolution of dense gas for three representative simulations: R10 (black), R10-kinf-0.1 (light blue) and R10-kinf-var-dtlim (magenta).  {\it Top panel:} The density of the densest gas cell within the two most massive progenitors of the final target halo that merge at $z=1.8$.  This cell has the greatest likelihood of forming a star particle.  The density threshold for star formation (which is fixed in comoving units) is shown as a dashed black line. {\it Bottom panel:} Total mass of gas at densities above half the threshold density for star formation in the simulation volume.}
\label{fig:rhomax_comp}
\end{figure}

This stark difference in the star formation rate can be explained by the evolution of dense gas within halos as shown in Figure~\ref{fig:rhomax_comp}.  In the runs which lack momentum input (e.g. R10), the maximum gas density in the main halo continues to rise significantly above the star-formation density threshold after the onset of star formation and feedback.  In simulations with some amount of kinetic energy injection, however, the maximum gas density in these halos appears to hew closely to the density threshold for star formation chosen in our models or to fall far below it.  This trend indicates that supernova feedback closely regulates star formation in this model, i.e. once gas becomes dense enough to form stars, the momentum feedback produced is able to prevent further collapse and thereby maintains the density of gas at the star-forming threshold.   We also see that there is some difference between the fast (R10-kinf-0) and slow (R10) injection of thermal energy: the slow injection of thermal energy in R10 appears to be effective in destroying self-shielded clumps of dense gas after reionization, unlike the quick injection of thermal energy in R10-kinf-0, where dense clumps survive to late times and continue to form stars.

If we look at the total amount of gas at or above the star formation density threshold, also shown in Figure \ref{fig:rhomax_comp}, we find that R10 and R10-kinf-0.1 have comparable amounts, and in fact R10-kinf-0.1 maintains somewhat more mass at the density threshold for a longer time (but, as shown in the top panel of Figure \ref{fig:rhomax_comp}, not much above this threshold).  Despite this, the overall efficiency of star formation is significantly lower in R10-kinf-0.1.  The self-regulating behavior of the kinetic feedback results in a larger reservoir of dense gas during the halo's star-forming epoch in R10-kinf-0.1, while in R10 with its purely thermal feedback, this excess of dense gas will cool to densities above the star forming threshold and be converted to star particles.

We note that the self-regulating behavior in R10-kinf-0.3 (and the other momentum-based feedback runs) has the result that the calculation does not reach the very highest levels of resolution allowed in the simulation because the refinement of the cartesian mesh on the highest levels is primarily triggered by gas density, which is suppressed.  In all three simulations, the first star particles form at $z > 10$ on AMR level 9, where the resolution is less than 8 physical pc.  In R10 and R10-kinf-0, the AMR hierarchy quickly reaches the maximum allowed level of AMR (level 12, which has eight times better spatial resolution).  However, in R10-kinf-0.3, the AMR hierarchy does not descend below level 10, which has a physical resolution of 8.6 pc at redshift 4.  

Simulations R10-kinf-var and R10-kinf-var-dtlim use the variable kinetic injection fraction scaling described in Section \ref{sec:kinf_var}.  These runs attempt to adjust the amount of kinetic energy injected based on the physical resolution of the grid.  The result of this approach is a slightly elevated star formation history, however, the self-regulation of gas density appears to be similar to the constant $f_{\rm{kin}}$ runs.  Despite this elevated star formation rate, the amount of dense gas in simulation R10-kinf-var-dtlim during the star-forming epoch is smaller than in for example R10-kinf-0.1, a simulation with a constant fraction of kinetic energy.  It appears that the effect of piling up gas just below the star formation density threshold, as happens in R10-kinf-0.1, in which the momentum in supernova affected gas is likely overestimated, does not occur, or does not occur to the same degree.  We found that adjusting the timestep resolution with this scheme also impacted the star formation history, which we discuss in Section \ref{sec:dtlim}.

\begin{figure*}
\centering
\includegraphics[scale=0.7] {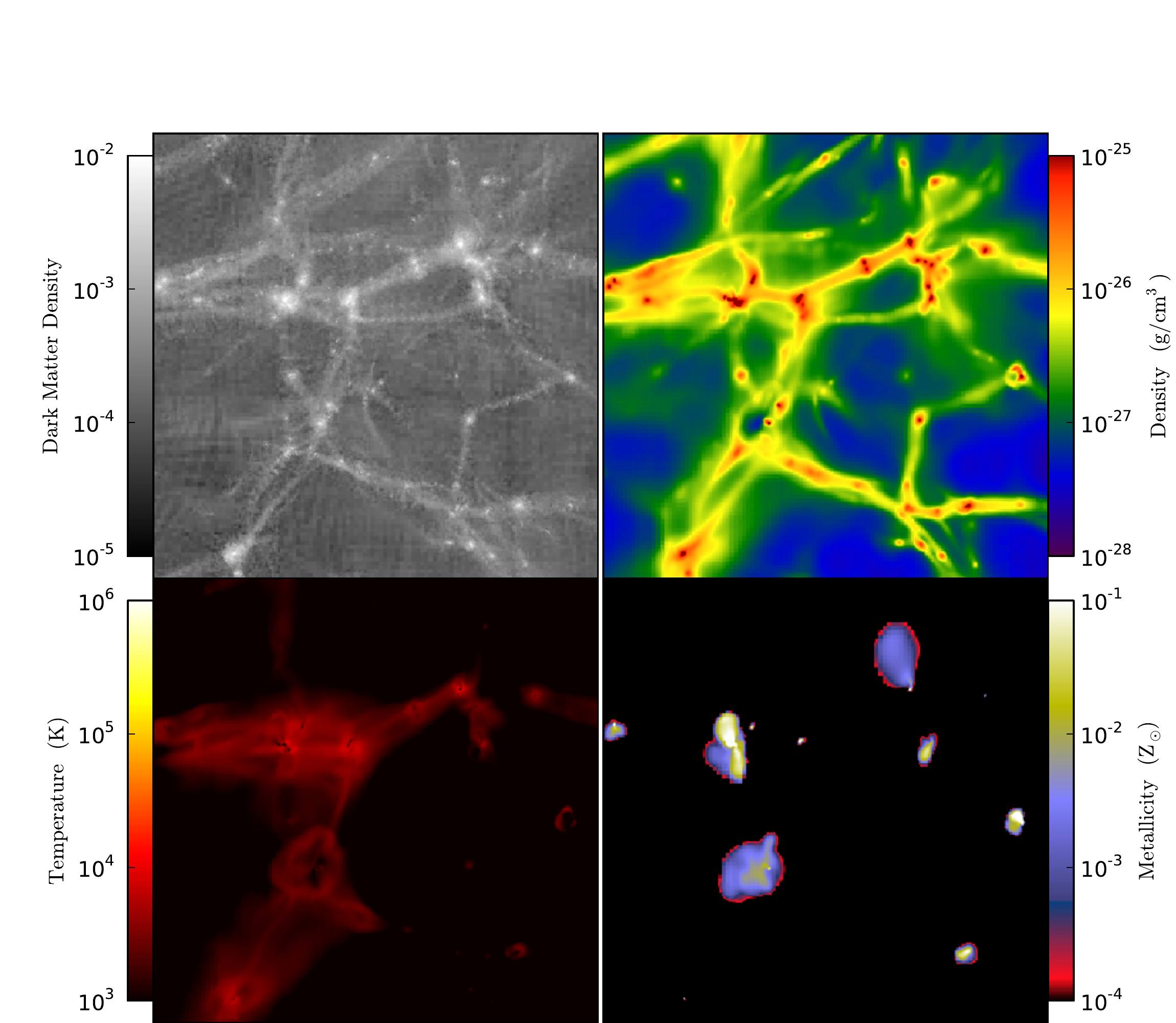}
\caption[Visualization of halo progenitors in R10-kinf-0]{Projections of dark matter density (top left), gas density (top right), gas temperature (bottom left) and gas metallicity (bottom right) at redshift 7.04 in simulation R10-kinf-0.  The volume projected is 300 comoving kpc, which is 37 kpc at this redshift.  This volume encloses all the star forming progenitor halos of the final target halo.  All the projections except for the dark matter density projection are weighted by gas density.  The dark matter density projection is unweighted. }
\label{fig:kinf0_image}
\end{figure*}

\begin{figure*}
\centering
\includegraphics[scale=0.7] {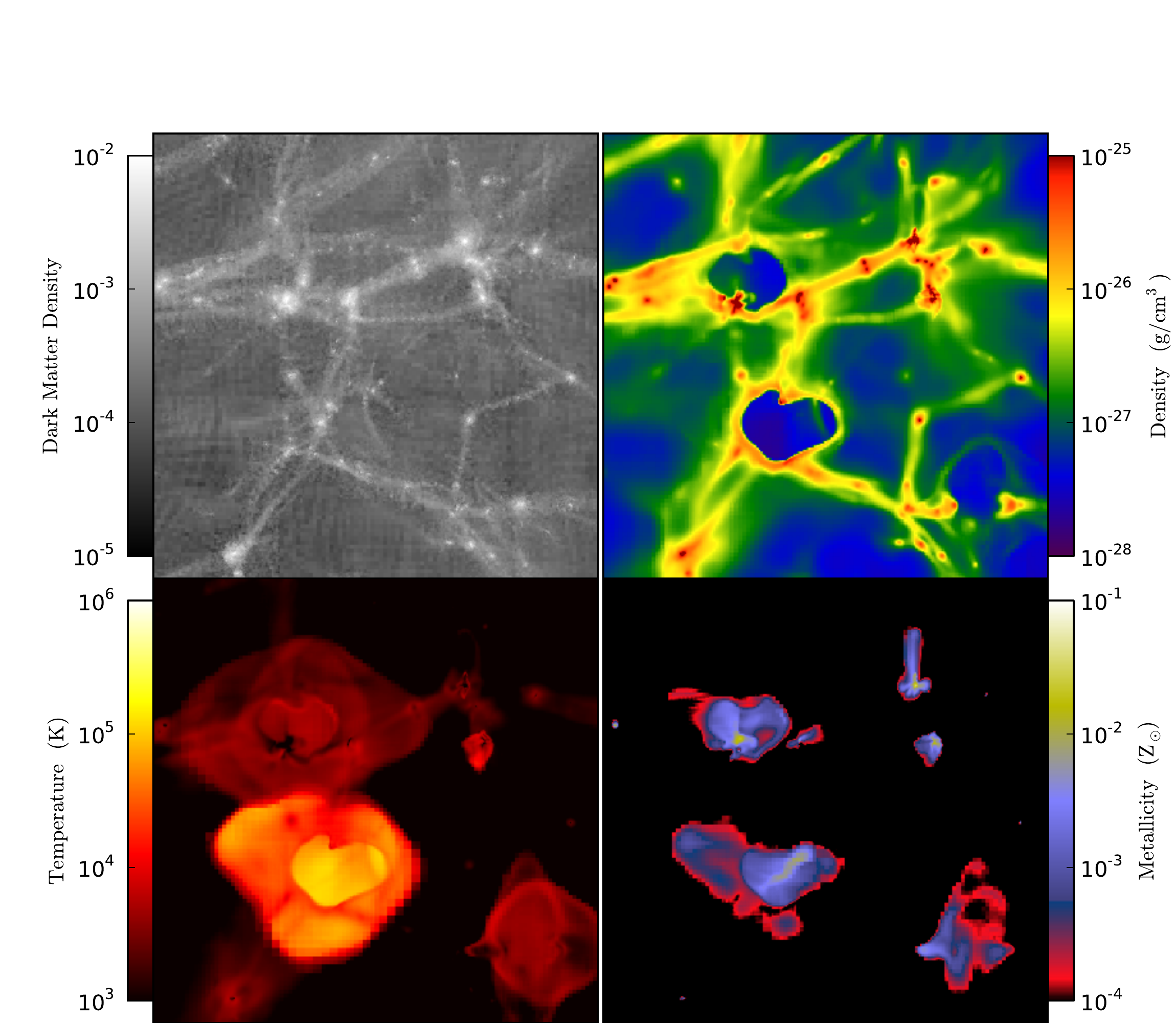}
\caption[Visualization of halo progenitors in R10-kinf-0.3]{Projections of simulation R10-kinf-0.3 at redshift $z=7.04$.  The quantities projected are the same as in Figure \ref{fig:kinf0_image}.}
\label{fig:kinf0.3_image}
\end{figure*}

The effect of different supernova feedback models manifests itself on larger scales as well.  Figures \ref{fig:kinf0_image} and \ref{fig:kinf0.3_image} show the environment around the population of star forming progenitor halos of the target halo just prior to the onset of reionization at $z=7.04$.  The global star formation rate in R10-kinf-0.3 is over an order of magnitude lower than it is in R10-kinf-0, but despite this, the supernova driven winds in R10-kinf-0.3 create large-scale, rarified bubbles of metal-enriched, $10^5$ K gas, unlike the much weaker winds driven in R10-kinf-0.

We have demonstrated that choosing a non-zero kinetic injection fraction for the feedback model can have a dramatic impact on the star formation rate of our simulated halo.  Given the results of Section \ref{sec:idealized_tests}, this is not surprising.  At the resolution of these simulations, and for the densities in which supernova feedback occurs in our model, we would expect the resolution timescale of supernova affected regions to be comparable to or to exceed the snow-plow timescale.  This means that radiative cooling dominates the energetic evolution before the kinetic and thermal energies have a chance to equilibrate.  This likely leads to a difference in the level of energy retained in supernova-heated gas between the purely thermal feedback models and the models with kinetic energy injection.  The result, as we have seen, is a more diffuse ISM in models with kinetic energy injection, and therefore a lower star formation rate.

What is perhaps more surprising is that only a very small amount of kinetic energy injection is needed to produce this effect.  All the simulations with some injection of kinetic energy behave very similarly.  Even the model with just 1\% or less of the supernova energy injected in kinetic form is much more similar to the model with 30\% injected kinetic energy than the model with 0\%.  As we saw in the idealized test simulations in Section \ref{sec:testing_setup}, the effect of kinetic energy injection on the ISM followed a smooth continuous trend with $f_{\rm{kin}}$ down to even the smallest values of $f_{\rm{kin}}$ tested.  The purely thermal tests were typically outliers of varying degree, depending on a multitude of physical and numerical factors.  In this example application that includes many star particles, on grids of varying density, resolution, metallicity and ionization state, this difference appears to be amplified, such that the purely thermal tests produce very different behavior.

As we have discussed, the density of the supernova affected region is crucial to the energetic evolution of the surrounding ISM.  In these simulations, we may be injecting energy into an unphysically dense medium, since we add supernova energy into the surrounding gas immediately after a star particle's creation where the gas density is high ($>$ 10 - 100 cm$^{-3}$), as set by the star formation algorithm.  Observations of supernova remnants in the Milky Way indicate that supernovae may occur in ISM with much lower densities ($\sim$ 1 cm$^{-3}$) \citep{berkhuijsen86}.  Type II supernovae are products of massive stars which produce large amounts of UV radiation and stellar winds prior to the supernova event.  The amount of energy produced in this phase is comparable to the amount of energy produced during the supernova \citep{agertz13} and may generate a rarified HII region surrounding the star.  We have not attempted to simulate this phase of stellar feedback although it may well be important in regulating the density of the ISM around supernova events.  On the other hand, the mass of material shed by a massive star during a red supergiant phase may raise the density of the surrounding ISM resulting in a higher background density \citep{vanveelen09}.  It is also interesting to note that the densities surrounding supernova remnants in the massive star forming region of the LMC 30 Doradus exceed the commonly found value in the Milky Way of 1 \cm\ \citep{smith04}.  

A lower background density for supernova events would reduce the importance of $f_{\rm{kin}}$ in our model because the snow-plow timescale would exceed the resolution timescale at the resolution of our galaxy model.
In addition, our resolution in feedback zones is not constrained to occur at the maximum allowed resolution in the simulation.  Rather, it is determined by the local gas and dark matter density.  In R10-kinf-0.3, the cell width does not drop below 43 comoving pc, which at $z=8$ corresponds to a cell width of more than 4 pc.  Based on Figure \ref{fig:multires_multiden}, if feedback at this redshift was constrained to occur on the finest level of allowed refinement, which has a factor of 4 better resolution, we may expect different behavior of supernova heated gas that is also less sensitive to the choice of $f_{\rm{kin}}$.  Most of the star formation and feedback in R10-kinf-0 does occur on grids at the maximum allowed resolution.  The reason for this difference, however, is that gas reaches much higher densities in R10-kinf-0 and therefore the resolution needed to reach a state where $f_{\rm{kin}}$ does not matter is higher. 

\subsection{Metal enrichment}
\label{sec:metals}

\begin{figure*}
\centering
\includegraphics[scale=1.0] {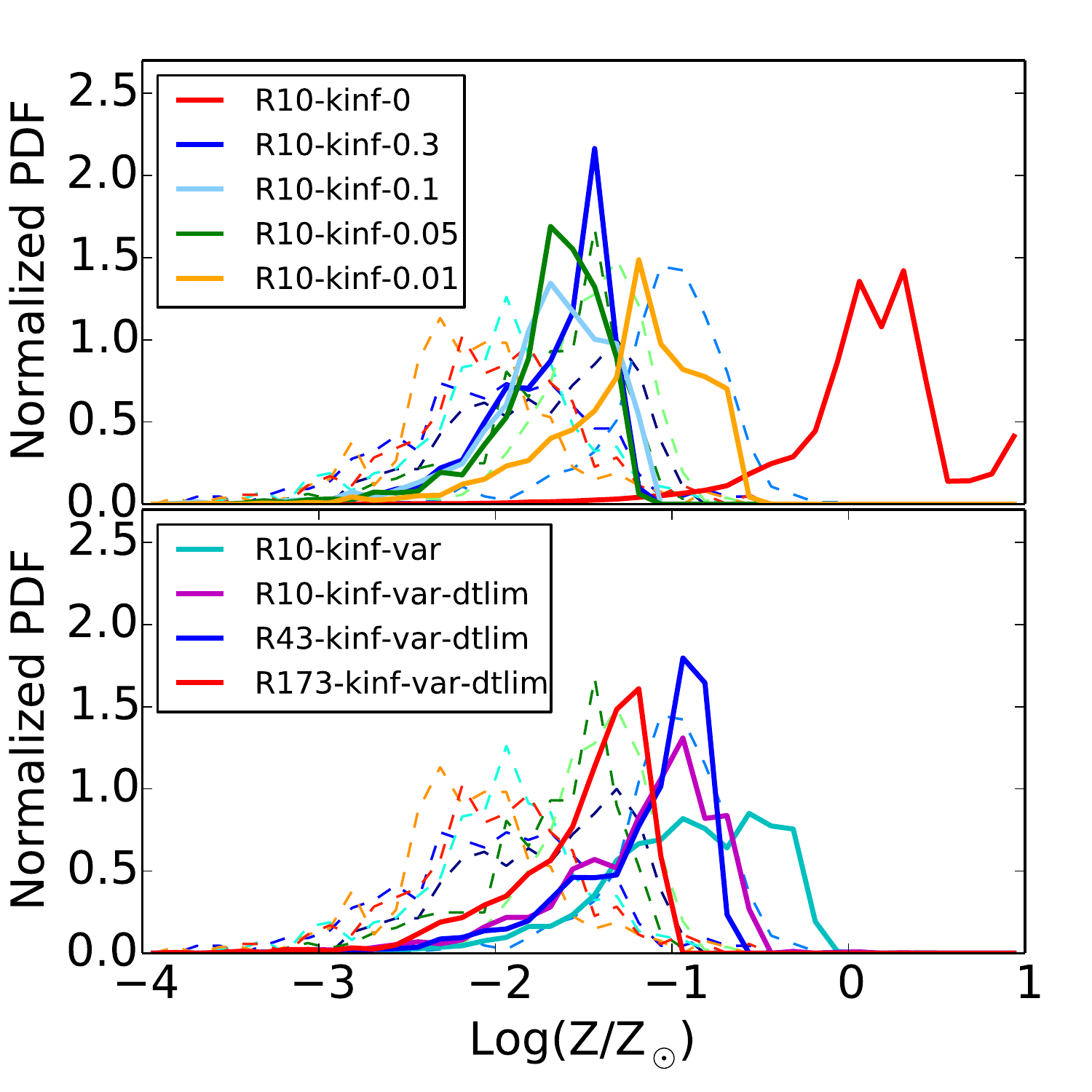}
\caption[]{Normalized distributions of stellar metallicity (assuming [Fe/H] = log(Z/\Zsun)) for observed dSphs of the Milky Way \citep[dashed lines]{kirby11a} compared to the metal fraction distributions within our simulated halos.  The simulated distributions include all star particles that end the simulation within $r_{200}$ of the halo center.  The top panel shows models with constant kinetic injection fractions and the bottom panel shows simulations with the resolution dependent scheme for kinetic energy injection.  The distributions plotted for R10-kinf-0 and R10-kinf-var, which were not run to $z=0$, show the metallicity distribution for star particles within the main halo at their ending redshift.}
\label{fig:obs_dist}
\end{figure*}

As seen in Figures \ref{fig:kinf0_image} and \ref{fig:kinf0.3_image}, the galactic winds produced in R10-kinf-0.3 (and all runs with momentum injection) are qualitatively more effective in enriching the IGM.  This effect can also be seen in the metals left behind in the halo's stellar component.  Table~\ref{tab} shows that the mean metallicity of the stellar population produced in models with kinetic feedback are much lower than in R10-kinf-0.

Figure \ref{fig:obs_dist} compares the metallicity distributions of the stellar populations in all our galaxy simulations run with the new supernova feedback model to metallicity distributions observed for dSphs in the Milky Way \citep{kirby11a}.  It is clear that the distribution found in R10-kinf-0 is inconsistent with observed systems due to its high mean, which is close to solar, while the distributions found in the models with momentum input are much closer to those found in observed systems.  For this comparison, we assume that [Fe/H] = Log(Z/\Zsun), where Z is the simulated mean metallicity field.

All simulations with kinetic energy input above 1\% appear to converge to a similar overall metallicity profile in their stellar populations.  Increasing the kinetic injection fraction from as little as 0.05 to as much as 0.3 does not result in a substantially increased loss of metals.  The model with 1\% kinetic energy begins to deviate from this common profile and shifts to higher metallicities, but still results in no star particles above solar metallicity, unlike R10-kinf-0.

As we have discussed, the variable kinetic energy injection scheme typically results in $f_{kin}<0.01$.  The metallicity distributions found in these runs are somewhat more enriched than the 1\% case, but not substantially so.  These runs also have a somewhat broader distributions.  We will further discuss these differences in Section \ref{sec:dtlim}.

The simulations conducted with the feedback model presented here still result in metallicities too high for their stellar masses.  All models except R10-kinf-var have stellar masses approximately less that $2\times 10^5$ \Msun.  Assuming that for an old population 2 \Msun\ of stellar mass produces 1 \Lsun\ in the V band \citep{kruijssen09} and given the luminosity-metallicity relation found for Local Group dwarfs \citep{kirby11a}, the mean stellar metallicity for these systems should be less than 2 dex in [Fe/H].  The conversion of our simulated metal fractions to the observed quantity [Fe/H] is not a straightforward one.  The metallicities in our simulations represent a mean over several chemical species, not just an iron abundance, which is the quantity the observational data measure.  Iron is produced primarily in Type Ia supernovae, so it is possible that stars formed in post-Type II supernova ejecta will be poorer in iron and richer in other elements and therefore, our calculation of [Fe/H] from our simulations is too high.  However, we are more interested the differences between metal fractions between models and the distribution of metal fractions within modes, so we choose not to address this issue in detail. 

As we have discussed, the kinetic feedback energy appears to closely regulate the density of gas in progenitor halos, keeping it fixed at the density threshold for star formation.  In addition, the metallicity of dense gas remains constant once it has enriched to an equilibrium level (not shown).  In our simulations, this results in a roughly constant star formation rate at a roughly constant stellar metallically and as a consequence, a large population of star particles is created at the equilibrium gas metallicity.  The spread in stellar metallicities in the variable kinetic fraction simulations is greater, however, they are still characterized by a low-metallicity tail and a sharper high-metallicity drop-off.  There are dwarfs that do exhibit distributions of this type, however, many do not.  In fact, observed systems of comparable stellar mass to our observed halos exhibit a diversity in stellar metallicity distributions \citep{kirby11a}.

A very likely reason for this difference is that the supernova feedback model we use produces a single, prompt injection of energy and metals, mimicking Type II supernovae, rather than a more comprehensive model that includes multiple sources of feedback occurring on different timescales and with different energies and chemistry \citep{agertz13}.  It is possible that this single source of stellar feedback naturally results in a sharp cutoff at high metallicity, especially in our case where feedback closely regulates gas density and metallicity.  

The simulation with the largest dispersion in stellar metallicities is R10, in which there is a gradual injection of metals and energy by star particles \citep{cenostriker92}.  This model creates a large population of low metallicity star particles.  It was originally intended for more massive star particles that release feedback energy from multiple sources from a well sampled stellar population.  While the application of this model to star particles of 100 \Msun\ is unphysical in many ways, it demonstrates that some ejection of metals through stellar winds in the pre-supernova stage may play a role in producing the broad distribution of stellar metallicities, especially at the low metallicity end.  

\subsection{Resolution Effects}
\subsubsection{Time resolution}
\label{sec:dtlim}

In Sections \ref{sec:sfr} and \ref{sec:metals}, we have shown the effects of both a constant kinetic injection fraction and the resolution-dependent kinetic injection scheme on star formation in this halo.  This scheme depends also on gas density and metallicity, such that in gas with shorter cooling times (i.e. higher density gas and/or higher metallicity gas) has higher kinetic injection fractions.  However, as resolution increases, this scaling assumes more of the supernova remnant's energy has been thermalized and therefore, a smaller $f_{\rm kin}$ is used.  
We found in R10-kinf-var and R10-kinf-var-dtlim that typically $R_{\rm{PDS}} < 4.5 \Delta x$ and therefore a small amount of kinetic energy was injected.  Usually, this value was less than 1\%.  In denser gas, the adaptive refinement increases resolution, but not sufficiently to ensure $R_{\rm{PDS}} < 4.5 \Delta x$.
 
The two galaxy tests using the variable $f_{\rm kin}$ method differ in how they limit time steps.  All the simulations presented here use the default Courant-limited time stepping implemented in Enzo \citep{bryan14}.  In addition, R10-kinf-var-dtlim uses $t_{\rm{PDS}}$ of typical star-forming gas to limit the time step just before the feedback event as described in Section \ref{sec:galaxy_methods}.  We found that this improvement in the time-step resolution, independent of an increase in mass or spatial resolution, increased the effectiveness of feedback.  This is likely an effect of better capturing the initial, pressure-driven flows from feedback-shocked gas.  

The differences between R10-kinf-var and R10-kinf-var-dtlim are twofold: one, the stellar populations in R10-kinf-var enrich to a slightly higher level; and two, star formation in small progenitor halos is not as effectively quenched in R10-kinf-var.  Figure \ref{fig:sfr_comp} shows that the star formation history of the main halo, which has assembled most of its dark matter by $z = 1.14$, is similar between the two runs with only a small delay in quenching time in R10-kinf-var.  The main differences arise from a smaller, satellite halo in R10-kinf-var that continues to form stars efficiently after the main halo quenches and keeps the total star formation rate of the simulation box elevated until the end of the simulation.  
The survival of low-level star formation likely results from low levels of energy injection whose ineffectiveness is promoted by the lower time resolution used to capture pressure-driven flows. 

In many respects, R10-kinf-var-dtlim is very similar in behavior to R10-kinf-0.01, which is a simulation that injects a constant fraction of kinetic energy of 1\% in all feedback events.  The kinetic fractions in the two simulations that use the variable scheme are typically no more than 1\% and in fact, are very often less than 0.1\%, depending on the resolution and gas properties.  Winds in this suite of simulations are generated by two effects: one, the direct injection of kinetic energy that adds bulk velocity to cells; and two, flows driven by pressure gradients in the gas.  In simulations with low levels of kinetic energy injection like R10-kinf-0.01 and R10-kinf-var-dtlim, the second effect has a greater importance and we see some convergence in properties between these runs.  

\subsubsection{Spatial Resolution}
\label{sec:res}

\begin{figure}
\centering
\includegraphics[scale=0.46]{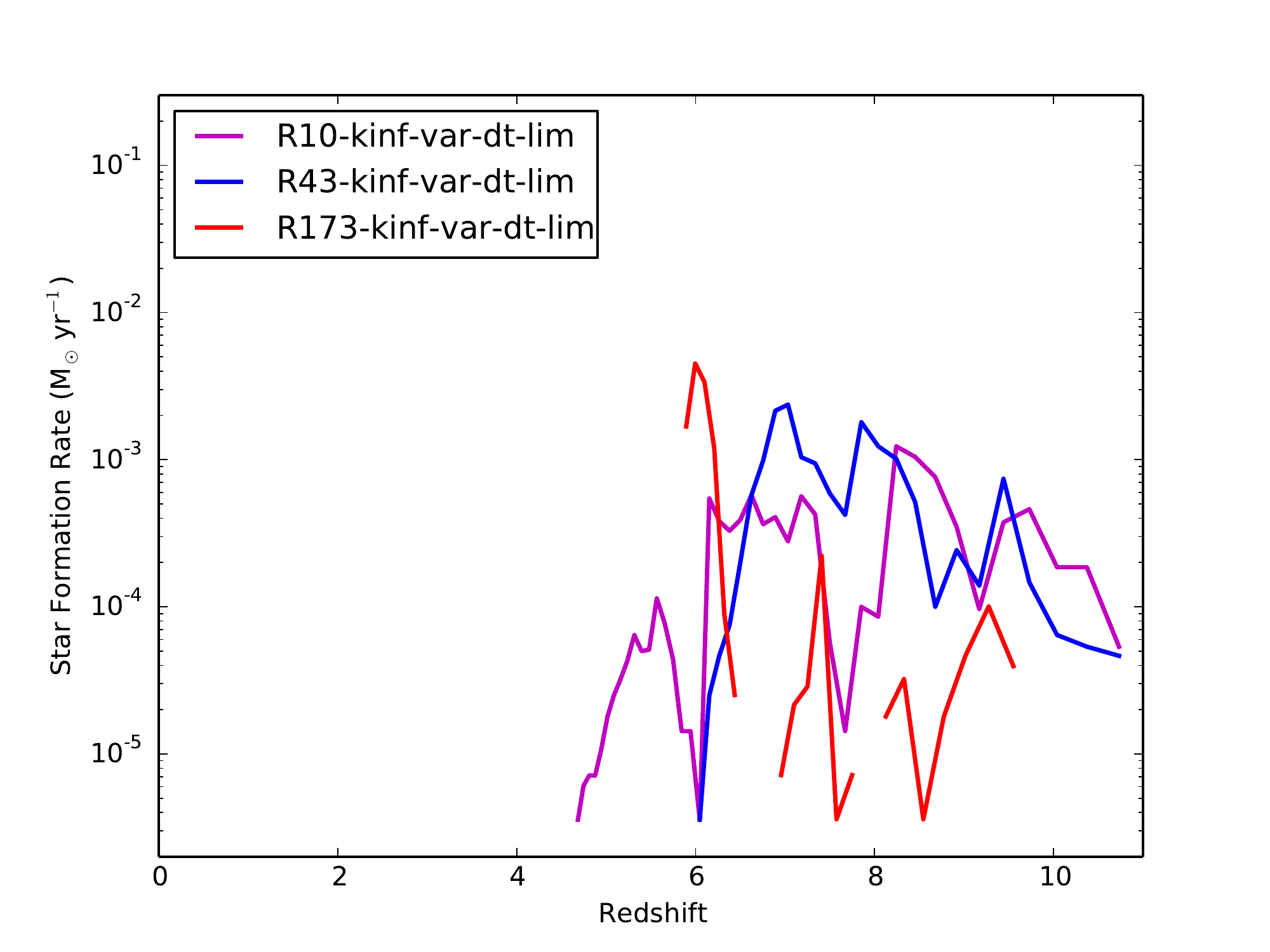}
\caption{Star formation rates of the target halo in simulations with varying minimum allowed cell width of 10.8 pc, 43 pc and 173 pc as described in Section \ref{sec:res} and Table \ref{tab}.  As in Figure \ref{fig:sfr_comp}, solid lines show the cumulative star formation history of all star particles that end the simulation within $r_{200}$.}
\label{fig:sfr_res}
\end{figure}

In all our cosmological zoom-in simulations, the AMR algorithm, which allows the simulation to reach higher resolutions than otherwise allowed in static grid simulations, means that in a single simulation hydrodynamics is solved with varying resolutions at a range of densities interesting for star formation.  Therefore, the resolution-dependent scaling for $f_{\rm{kin}}$ results in different kinetic injection fraction not just between simulations of overall different resolutions, but also within a single simulation.  As is typically the case in AMR simulations, we allow refinement in such a way to maintain a cells with a constant mass in a way that is analogous to Lagrangian methods.  Therefore, at each AMR level, the maximum density that can be reached (before refinement is triggered), is proportional to $2^{3l} (z + 1)^3 \overline{\rho}$, where $l$ is the AMR level (allowed up to 12) and $\overline{\rho}$ is the mean density of the universe at $z= 0$.  Therefore, the variable scaling of $f_{\rm{kin}}$ used in some of our cosmological simulations in principle also depends on redshift and AMR level through its density dependence.

To better understand how spatial resolution allowed in the hydrodynamics calculation impacts our model when run with AMR in a realistic galactic setting, we have also undertaken two simulations R43-kinf-var-dtlim and R173-kinf-var-dtlim where the minimum cell width is limited to 43 pc and 173 pc respectively.  Table \ref{tab} summarizes some of the properties of the target halo in these simulations and Figure \ref{fig:sfr_res} presents the star formation histories within the main halos in these simulations.

We have altered nothing else in these simulations apart from the maximum allowed level of AMR.  The maximum allowed mesh resolution in simulation R173-kinf-var-dtlim is the lowest resolution where densities above the star-formation threshold can be reached.  Star-forming gas is resolved in this simulation, but not well, and the result is that star formation proceeds at a relatively low-level (compared to the higher resolution simulations), is rather bursty and is dominated by fewer star-forming progenitor halos.  These differences should be understood as due primarily to the inefficiency of star formation rather than to an effect of the feedback model.

The simulation R43-kinf-var-dtlim, however, does have sufficient resolution to resolve gas at densities more than an order of magnitude above the star-forming threshold; the behavior in this simulation is likely due more to the feedback model as implemented under different AMR conditions.  However, we see that overall, the properties and evolution of this simulation are remarkably similar to R10-kinf-var-dtlim.  This similarity could have been predicted from the fact that R10-kinf-var-dtlim has difficulty reaching the maximum allowed AMR level and in fact, the deepest level in the hierarchy that R10-kinf-var-dtlim reliably reaches is only one level more than the maximum of 10 levels allowed in R43-kinf-var-dtlim.  This contrasts with simulations of this same halo run at this resolution with purely thermal feedback as discussed in \citet{simpson13}.  In that simulation, R43, limiting the spatial resolution resolution to this level did have a significant effect on the overall evolution of star formation.  Thermal feedback in that simulation was much less effective at quenching star formation post-reionization due to the fact that the motions of supernova affected gas were controlled primarily by the specific heating rate from the supernova feedback and that rate is lower at lower grid resolutions, the classic `over-cooling problem.'

\section{Discussion}
\label{sec:discussion}

The model we have described and tested introduces a new tuneable parameter, $f_{\rm{kin}}$, the fraction of injected kinetic energy.  The model also has (like most star-formation algorithms) parameters that control the total amount of mass, metals, and energy to return to the gas, as well as the timescale over which to inject it.  In principle these other parameters can be determined in reference to the IMF and how it is sampled.  These choices are not trivial and have the potential to affect results, however in this study, we have deliberately chosen to fix those parameters in order to focus on $f_{\rm{kin}}$. 

The parameter $f_{\rm{kin}}$ is a fundamentally different type of parameter as it depends on the ability of the hydrodynamical solver to capture the state of shocked gas and depends primarily on resolution.  The resolution dependent behavior is a function of the background density and in section~\ref{sec:kinf_var}, we presented a density-dependent scaling for $f_{\rm{kin}}$, which also accounts for metallicity.  This model is a promising start, but it is not the final word -- a more complete model might build on the work presented here to prescribe both $f_{\rm{kin}}$ and the energy available based on the appropriate SN phase (free-expansion, Sedov, or snow-plow) and the resolution.  Nevertheless, even a constant value of $f_{\rm kin}$ may capture much of the key phenomenology, as we saw in Section~\ref{sec:galaxy_simulation} that simulations with any momentum input had very similar behavior, while simulations with zero kinetic energy were quite different.

The idealized test simulations we have presented can also be used to provide a guide for choosing simulation resolution.
For example, Figure~\ref{fig:final_efrac} shows that at all background densities tested, resolutions of 1-2 pc result in evolution that is largely independent of $f_{\rm{kin}}$ -- a similar resolution requirement is also found by turbulent, high-resolution ($\sim$ 1 pc) ISM simulations \citep[e.g.,][]{joung06}.
The galaxy simulations we present here already achieve resolutions of this order at high redshift.  Simulations with an overall higher resolution or a more creative adaptive refinement scheme that triggers refinement around feedback producing particles would ensure high resolution is reached around feedback events at all times.  The dependence on $f_{\rm{kin}}$ may be less in such simulations.  

In addition, an important point that should be emphasized is that the resolution dependent behavior of the model is much weaker in lower density media.  This has important consequences when considering the numerous physical effects we neglect in our galaxy model.  For example, coupling this model with a model for radiation feedback from young stars may serve to lower the resolution requirements of the supernova feedback scheme by lowering the density of gas surrounding supernovae events \citep[e.g.][]{rosdahl15}.  Moreover, even without a radiation feedback scheme, simply allowing star particles to drift from their birth sites, which are necessarily in the highest density gas cells, may by itself produce a similar effect \citep{renaud13}.  In this study, we are in a sense injecting feedback events in the least hospitable conditions for thermal energy injection.  This in effect maximizes the impact of kinetic energy injection.  Ideally, future work would include a more complete model, simulated at higher resolution, where this effect would instead be minimized.

\section{Summary and Conclusions}

In this paper, we have described a novel scheme for adding energetic feedback to a grid-based hydrodynamics code for use in galaxy simulations.  The new aspects of this method include: (i) a compact feedback stencil which naturally builds on particle-grid interaction methods (such as the cloud-in-cell method) in a relatively isotropic fashion; (ii) the use of the existing solution on the grid to set a specific amount of kinetic and thermal energy to be added to the grid; and (iii) a net vector momentum input of  zero.   

The free parameter in the model is $f_{\rm kin}$, the fraction of the energy injected in kinetic form.  We have extensively tested the method in homogeneous media, first confirming that, in the absence of radiative cooling, we reproduce the expected Sedov-Taylor solution.  Then, we add cooling and carry out a large number of (homogeneous) test calculations in which we vary: (i) the kinetic to thermal injection fraction; (ii) the density of the background medium; and (iii) the resolution.  We explore the results, characterizing them primarily by the amount of energy remaining after a fixed amount of time (typically 0.1 Myr) and show how they can be understood simply through a set of three timescales that describe how the solution is affected by (in turn) purely numerical artifacts, the approach to the self-similar Sedov-Taylor solution, and radiative cooling.  

Based on this analysis, we present an analytic scheme for choosing an appropriate $f_{\rm kin} (\Delta x, \rho, Z)$ given the resolution, density and metallicity of simulated gas, so that the resulting energy (after 0.1 Myr) closely matches the results from high-resolution ($\Delta x$ = 1 pc) simulations of blast-wave evolution in a homogeneous, radiative medium.

Finally, we used our feedback scheme to model the formation of a low-mass dwarf system in a cosmological setting.  We found that constant kinetic injection fractions between 1\% and 30\% result in qualitatively similar results that differ greatly from models that include only thermal energy injection.  Kinetic feedback results in an order of magnitude suppression of the total stellar mass and approximately a 1 dex drop in the average stellar metallicity.  The variable scheme for kinetic energy injection that we tested is somewhat less effective in dispersing metals and suppressing star formation, however, the results still differ drastically from the zero kinetic energy runs.

The method we have presented is, in principle, scale independent and applications of the numerical scheme can be adapted for subgrid feedback recipes for use in lower-resolution calculations in future work.  In the high-resolution ($\Delta x <$ 10 pc) application that we explored in this study, we have used a simplistic model for the sources of stellar feedback, restricting the model to Type II supernovae.  Future work that includes a more sophisticated sampling of supernovae events from the stellar particles (e.g. Type Ia supernovae and high-velocity OB stars) as well as models for UV heating from young stars and winds from evolved stars will likely produce models with higher fidelity to observations.  The basic prescription for adding kinetic energy could also be extended to include feedback from AGN.

 \section{Acknowledgements}
The authors would like to thank the anonymous reviewer for thoughtful comments and suggestions that helped improve the paper.  This work was supported by NASA grant NNX12AH41G.  GLB also acknowledges support from grants NSF-1008134, NSF-1210890, and NSF-1312888.   CMS acknowledges support by the European Research Council under ERC-StG grant EXAGAL-308037.  Computational resources from NSF XSEDE, SDSC, TACC and Columbia University's shared research compute cluster were used to perform this work.  This work was supported in part by the National Science Foundation under Grant No. PHYS-1066293 and the hospitality of the Aspen Center for Physics.  The authors would also like to thank Mordecai Mac Low, Volker Springel and Federico Marinacci for their valuable comments on our draft and Marla Geha, Mary Putman and John Wise for valuable comments in the development of this work.
 

\end{document}